\documentclass[11pt]{article}

\usepackage[colorlinks, linkcolor=blue, urlcolor=blue, citecolor=blue]{hyperref}           
\usepackage{color}
\usepackage{graphicx,subfigure,amsmath,amssymb,amsfonts,bm,epsfig,epsf,url,dsfont}
\usepackage{amsthm}

\usepackage[export]{adjustbox}

\newtheorem{thm}{Theorem}[section]

\newtheorem{proposition}[thm]{Proposition}
\newtheorem{remark}[thm]{Remark}

\usepackage{tikz}
\usepackage{bbm}
\usepackage[small,bf]{caption}
\usepackage{natbib}
\usepackage[top=1in,bottom=1in,left=1in,right=1in]{geometry}
\usepackage{fancybox}
\usepackage{algorithm}
\usepackage{verbatim}

\usepackage{mathtools}

\usepackage{scalerel,stackengine}
\stackMath
\newcommand\reallywidehat[1]{%
\savestack{\tmpbox}{\stretchto{%
  \scaleto{%
    \scalerel*[\widthof{\ensuremath{#1}}]{\kern-.6pt\bigwedge\kern-.6pt}%
    {\rule[-\textheight/2]{1ex}{\textheight}}%WIDTH-LIMITED BIG WEDGE
  }{\textheight}% 
}{0.5ex}}%
\stackon[1pt]{#1}{\tmpbox}%
}

\newcommand{\tr}{\mathrm{tr}}

\newcommand{\der}{\mathrm{d}}

\newcommand{\sT}{\mathsf{T}}

\newcommand{\F}{\text{F}} 
\newcommand{\R}{\mathbb{R}}

\newcommand{\E}{\mathbb{E}}

\DeclareMathOperator*{\argmax}{arg\,max}
\DeclareMathOperator*{\argmin}{arg\,min}
\def\P{{\mathbb P}}

\newcommand{\p}[1]{\left(#1\right)}
\newcommand{\pp}[1]{\left[#1\right]}
\newcommand{\ppp}[1]{\left\{#1\right\}}
\newcommand{\norm}[1]{\left\|#1\right\|}

\newcommand{\barV}{{V}}%
\renewcommand{\i}{\mathbf{i}}

\newcommand{\s}[1]{\mathsf{#1}}

\usepackage{xspace}
\usepackage[titletoc,toc]{appendix}

\numberwithin{equation}{section}

\allowdisplaybreaks

\begin{document}

\title{Bayesian Perspective for Orientation Determination in Cryo-EM with Application to Structural Heterogeneity Analysis}

\author{
    Sheng Xu\thanks{Corresponding author; email: \href{mailto:sxu21@princeton.edu}{sxu21@princeton.edu}} \footnote{These authors contributed equally.} \\ 
    \small Program in Applied and Computational Mathematics, Princeton University, Princeton, NJ 08544, USA
    \and 
    Amnon Balanov\footnotemark[2] \\ 
    \small School of Electrical Engineering, Tel Aviv University, Tel Aviv 69978, Israel~~
    \and 
        Amit Singer \\ 
    \small Program in Applied and Computational Mathematics and Department of Mathematics,\\
    \small Princeton University, Princeton, NJ 08544, USA
    \and 
    Tamir Bendory \\ 
    \small School of Electrical Engineering, Tel Aviv University, Tel Aviv 69978, Israel
}

\vspace*{-60pt}
    {\let\newpage\relax\maketitle}

\begin{abstract}
Accurate orientation estimation is a crucial component of 3D molecular structure reconstruction, both in single-particle cryo-electron microscopy (cryo-EM) and in the increasingly popular field of cryo-electron tomography (cryo-ET). The dominant approach, which involves searching for the orientation that maximizes cross-correlation relative to given templates, is sub-optimal, particularly under low signal-to-noise conditions. In this work, we propose a Bayesian framework for more accurate and flexible orientation estimation, with the minimum mean square error (MMSE) estimator serving as a key example. Through simulations, we demonstrate that the MMSE estimator consistently outperforms the cross-correlation-based method, especially in challenging low signal-to-noise scenarios, and we provide a theoretical framework that supports these improvements.

When incorporated into iterative refinement algorithms in the 3D reconstruction pipeline, the MMSE estimator markedly improves reconstruction accuracy, reduces model bias, and enhances robustness to the ``Einstein from Noise'' artifact. Crucially, we demonstrate that orientation estimation accuracy has a decisive effect on downstream structural heterogeneity analysis. In particular, integrating the MMSE-based pose estimator into frameworks for continuous heterogeneity recovery yields accuracy improvements approaching those obtained with ground-truth poses, establishing MMSE-based pose estimation as a key enabler of high-fidelity conformational landscape reconstruction. These findings indicate that the proposed Bayesian framework could substantially advance cryo-EM and cryo-ET by enhancing the accuracy, robustness, and reliability of 3D molecular structure reconstruction, thereby facilitating deeper insights into complex biological systems.

\end{abstract}
    
\section{Introduction}

\begin{figure}
    \centering
    \includegraphics[width=1.0\linewidth]{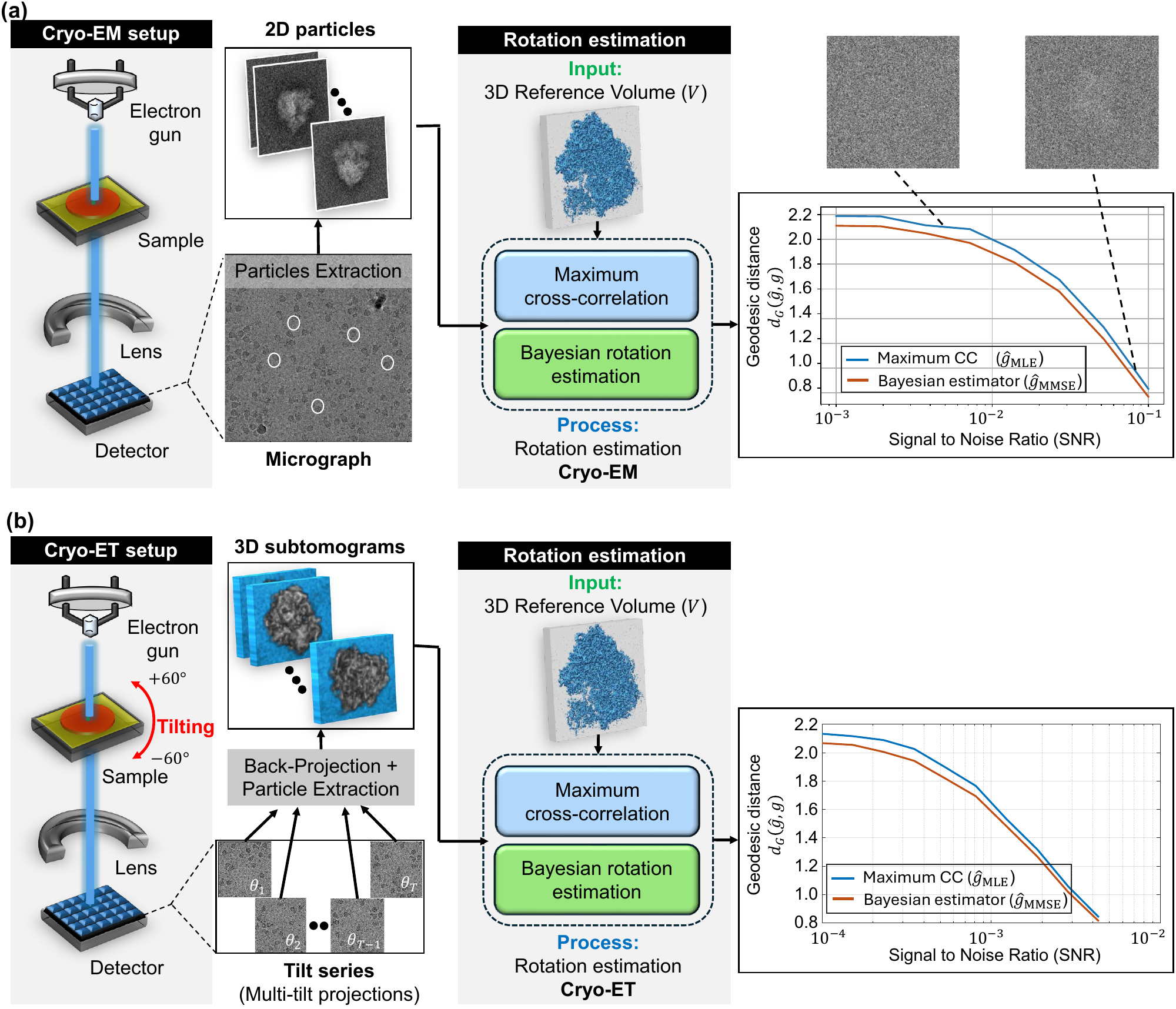}
    \caption{\textbf{Comparison between maximum cross-correlation ($\hat{g}_{\text{MLE}}$) and Bayesian ($\hat{g}_{\text{MMSE}}$) orientation estimators in cryo-electron microscopy (cryo-EM) and cryo-electron tomography (cryo-ET) applications.} The figure illustrates the general workflow in cryo-EM and cryo-ET techniques, highlighting the role of orientation estimation in each technique. Panel \textbf{(a)} illustrates the model \textit{with 2D projections} (single-particle cryo-EM model, \eqref{eq:cryoEM}), while panel \textbf{(b)} shows the model of the subtomogram averaging in cryo-ET, \eqref{eq:cryoET}).   
    \textbf{(a) Cryo-EM} involves imaging macromolecules embedded in a thin layer of vitreous ice using an electron beam in a transmission electron microscope (TEM). The process generates 2D projection images (micrographs) of particles in unknown 3D orientations. These 2D particles are then identified and extracted from the micrographs, forming the basis for subsequent steps of the macromolecule's 3D structure reconstruction.
    \textbf{(b) Cryo-ET} involves imaging a sample from multiple known tilt angles (typically from $-60^\circ$ to $+60^\circ$) to create 2D projections, which are then combined computationally to reconstruct 3D subtomograms. In this context, a subtomogram refers to a small volume containing an individual 3D particle. The subtomograms are extracted by a particle picker algorithm for further analysis.  
    \textbf{(a+b):} The rotation estimation problem involves determining the relative orientation of a noisy 2D particle (in cryo-EM) or a noisy 3D subtomogram (in cryo-ET) relative to a reference volume $V$. The reference volume structure used in both setups is identical and corresponds to the 80S ribosome~\cite{wong2014cryo}.
    Under high SNR conditions, both rotation estimators closely approximate the true relative rotation. However, as the SNR decreases, the estimation accuracy deteriorates. Importantly, across all SNR levels, the geodesic angular distance between the MMSE orientation estimator and the true rotation consistently remains lower than that of the MLE orientation estimator. 
    For \textbf{(a)}, the estimation was conducted using a grid size of $L=3000$ samples of the rotation group $\s{SO}(3)$, while for \textbf{(b)}, a grid size of $L=300$ was used. Each point in the two curve plots represents the average error computed over 3000 trials.}
    \label{fig:fig1} 
\end{figure}

Determining the precise three-dimensional (3D) orientation of biological molecules from their noisy two-dimensional (2D) projection images is a fundamental challenge in cryo-electron microscopy (cryo-EM) \cite{bai2015cryo,lyumkis2019challenges,bendory2020single}. This process, known as orientation estimation, is crucial for various cryo-EM applications, including 3D reconstruction algorithms \cite{scheres2012relion, punjani2017cryosparc}, heterogeneity analysis \cite{sorzano2019survey, toader2023methods, donnat2022deep,zhong2021cryodrgn}, and beyond~\cite{maeots2022structural}. For example, Figure \ref{fig:fig1}(a) illustrates the role of orientation estimation within the cryo-EM 3D reconstruction workflow.

In cryo-electron tomography (cryo-ET), orientation estimation presents an additional challenge in the form of subtomogram averaging. Notably, cryo-ET suffers from higher noise levels compared to single-particle cryo-EM due to the complex and heterogeneous nature of cellular samples and the challenges of capturing data from multiple angles within thicker specimens. Subtomogram averaging offers an effective approach to enhance the signal-to-noise ratio (SNR), ultimately resulting in the reconstruction of high-resolution structures. This technique often involves extracting multiple similar subtomograms containing the target protein complex or macromolecule from a large cryo-electron tomogram reconstructed from all available tilts (typically from $-60^\circ$ to $+60^\circ$), followed by aligning and averaging them \cite{zhang2019advances,watson2024advances}. Unlike traditional cryo-EM, this process typically aligns 3D structures directly, without the direct use of 2D projections (see Figure \ref{fig:fig1}(b) for an illustration).

Mathematically, the orientation estimation tasks in cryo-EM and cryo-ET are slightly different.
In the process of single-particle cryo-EM, which involves 2D tomographic projections, the mathematical model can be formulated as:
\begin{equation} \label{eq:cryoEM}
y = \Pi(g\circ V) + \varepsilon,
\end{equation}
where $y: \R^2\to\R$ is the observed 2D projection image, $V: \R^3\to\R$ is the underlying 3D molecular structure, $\Pi$ is the tomographic projection operator, $g$ is the unknown 3D rotation operator of interest, $\varepsilon$ represents measurement noise, and $g \circ V(x) \equiv V(g^{-1} x)$, representing a rotation acting on a volume $V$ with 3D coordinate $x$. 
Analogously, the mathematical model in cryo-ET subtomogram averaging, which aligns directly with the 3D structure, can be represented by:
\begin{equation} \label{eq:cryoET}
y = g \circ V + \varepsilon,
\end{equation}
where $y: \R^3\to\R$ is the observed 3D subtomogram, $V: \R^3\to\R$ is the underlying 3D molecular structure, $g$ is the unknown 3D rotation operator of interest, and $\varepsilon$ represents measurement noise. Then, the goal of orientation estimation is \textit{to find the ``best'' 3D rotation $g$ based on the 2D projection image (in the cryo-EM case~\eqref{eq:cryoEM}) or the 3D subtomogram (in the cryo-ET case~\eqref{eq:cryoET}) with respect to the 3D reference $V$. Namely, we aim to estimate the rotation $g$ given the sample $y$ and the 3D structure $V$.}

The actual mathematical models used in cryo-EM and cryo-ET are more intricate than those presented in \eqref{eq:cryoEM} and \eqref{eq:cryoET}, incorporating additional factors such as the contrast transfer function (CTF) and in-plane translations, as elaborated in Appendix \ref{appx:fullCryoEmModel}. Although the framework introduced in this work is capable of addressing the full cryo-EM model, we adopt the simplified models in \eqref{eq:cryoEM} and \eqref{eq:cryoET} to more effectively communicate the primary insights.

\paragraph{The gap.}
The common approach to estimating the rotation of an observation in the models above involves scanning through a pre-defined set of possible rotations and selecting the one that either maximizes the correlation or minimizes the distance to the given reference (2D projection image or 3D molecular structure, depending on the application). Typically, these metrics involve weighted correlations and distances, where the weights account for the noise characteristics \cite{scheres2012relion}.

From an estimation theory perspective, this process corresponds to the maximum likelihood estimator (MLE), which does not incorporate prior information about the \textit{rotation distribution}, i.e., the distribution of the 3D rotation $g$ appearing in models \eqref{eq:cryoEM} or \eqref{eq:cryoET}. A natural extension is the maximum a posteriori (MAP) estimator, which combines the data likelihood with a prior distribution on rotations. Notably, when a uniform prior is assumed, indicating that all orientations are equally likely a priori, the MAP estimator reduces to the MLE estimator~\cite{steven1993fundamentals}. A more rigorous treatment can be found in Section \ref{sec:mmseMapAndMLest}.

However, Bayesian theory provides a much deeper and richer statistical framework that leads to improvement: replacing the MLE or MAP estimator with the \textit{Bayes estimator}. The full potential of this estimator, that provides optimal accuracy according to a user-defined loss function and allows for integrating prior knowledge about \textit{rotation distribution}, remained untapped so far.

The term \textit{rotation distribution}, used interchangeably with orientation distribution in this paper, refers to the probabilistic law governing how orientations are distributed over all possible rotations in 3D space. In cryo-EM, this concept is critical because molecules are first suspended in a thin aqueous solution and subsequently vitrified into amorphous ice, at which point their orientations become fixed but remain uncontrolled. Ideally, all orientations would be equally likely, corresponding to a uniform distribution. In practice, however, molecules often adopt \textit{preferred orientations}, typically due to interactions with the air–water interface or other sample-preparation factors \cite{tan2017addressing,lyumkis2019challenges, li2021effect}. As a result, certain orientations are disproportionately represented, introducing systematic biases that can degrade reconstruction quality and complicate orientation estimation. Accounting for these deviations from uniformity is therefore essential for both methodological development and the interpretation of cryo-EM data.

Currently, the most practical way to obtain orientation estimates is through modules of reconstruction software such as \textsc{RELION}~\cite{scheres2012relion} and \textsc{cryoSPARC}~\cite{punjani2017cryosparc}. 
These packages are primarily designed for 3D structure reconstruction and therefore treat pose estimation as a latent-variable subproblem. 
In particular, during refinement they compute  posterior weights over a discrete set of candidate poses for each particle and use these weights for weighted averaging in the reconstruction step (a soft-assignment over poses). 
At the same time, for downstream usage and reporting, the metadata typically provides a \emph{single} orientation per particle, which is commonly the MLE estimator, that is, the maximizer of the pose posterior. Thus, while these pipelines do incorporate latent-variable pose weighting internally for reconstruction, they do not typically output a Bayesian rotation estimate for each particle.

In practice, researchers frequently use the reported per-particle orientation point estimates for downstream tasks such as heterogeneity analysis~\cite{gilles2025cryo, zhong2021cryodrgn} and structural validation~\cite{rosenthal2003optimal}. 
However, relying solely on MLE-based estimates can be suboptimal, especially in low-SNR regimes where MLE estimators are known to perform poorly. 
This work therefore explores opportunities to improve per-particle orientation estimation by using Bayes estimators that explicitly average over pose uncertainty that incorporate prior knowledge of the rotation distribution.

\paragraph{The Bayesian framework.} 
The Bayesian framework has become a powerful and widely adopted tool in cryo-EM, now recognized as the leading method for recovering 3D molecular structures \cite{scheres2012relion, punjani2017cryosparc, toader2023methods, gilles2025cryo}. It effectively addresses challenges like overfitting and parameter tuning while enhancing interpretability~\cite{scheres2012bayesian}. By explicitly modeling uncertainties, Bayesian methods enable more accurate and robust reconstructions of molecular structures, driving significant advances in both resolution and structural flexibility. While the majority of these methods focus on the task of structure reconstruction (e.g., see Section \ref{sec:poseDetrminantionVolumeReconstruction}) and aim to achieve the MAP estimator for the volume structure~\cite{scheres2012relion, punjani2017cryosparc}, the Bayesian framework offers broader possibilities for addressing other problems. 

 This framework enables the use of any loss function over the rotational group tailored to users’  requirements and accommodates a broad range of prior distributions, beyond the uniform distribution. While the Bayesian framework can be adapted to different loss functions, this work focuses on the mean-squared error loss, which is equivalent to the chordal distance between 3D rotations. The primary reason for using this loss function is that its corresponding Bayes estimator has a closed-form analytical solution, making it computationally efficient and easy to interpret. We denote this estimator as $\hat{g}_{\text{MMSE}}$, where MMSE stands for minimum mean square error. It is worth noting that for any given loss function and prior distribution, the Bayes estimator is optimal among all possible estimators, in the sense that it minimizes the posterior expected loss. Moreover, the computational cost of calculating our $\hat{g}_{\text{MMSE}}$ estimator is at the same scale as that of the commonly used MLE and MAP orientation estimators (See Section \ref{sec:numericalMethods} for more details). 

\paragraph{Applications.}

Orientation estimation is not only integral to 3D structure reconstruction but also underpins a range of downstream tasks. Below, we highlight several representative applications that demonstrate its broader methodological significance.

One prominent example is heterogeneity analysis, where the goal is to capture structural variability that may arise from differences in composition, discrete states, or continuous conformational changes. A substantial portion of state-of-the-art methods \cite{gilles2025cryo, zhong2021cryodrgn, levy2024end, luo2023opus, punjani2017cryosparc, punjani20213d} make the simplifying assumption that particle poses are already known and fixed, typically obtained from a consensus refinement procedure; these pose estimates then form the basis for subsequent modeling of structural variability. In this work, we consider continuous conformational heterogeneity and demonstrate in Section~\ref{sec:heterogeneityAnalysis} that the quality of orientation estimation has a direct impact on the fidelity of the recovered conformational landscapes.

Another important application is 3D volume alignment. Unlike the setting of~\eqref{eq:cryoET}, where an observation is aligned to a known reference volume, this task involves aligning two noisy volumes of the same structure with unknown relative rotations. This scenario arises, for instance, in the computation of the Fourier Shell Correlation (FSC) curve, a standard tool for estimating the spectral signal-to-noise ratio (SSNR) and resolution. As shown in Appendix~\ref{appx:alignmnetBetweenNoisyVolumes}, this problem can be expressed within the same statistical framework as~\eqref{eq:cryoET}, differing only by an additional noise term, which highlights its close methodological connection to orientation estimation.

A further noteworthy application concerns validation. A classic example is the method of tilt pairs introduced by Rosenthal and Henderson~\cite{rosenthal2003optimal}, in which a small subset of images (e.g., around ten) is collected at known relative tilt angles. While the absolute particle orientations remain unknown, the relative orientations between tilt pairs are specified by the experimental geometry. The degree of agreement between estimated and known relative rotations then provides a direct validation of orientation assignment accuracy. 

These examples are by no means exhaustive, but they underscore that orientation estimation is not merely a technical nuisance in reconstruction workflows, but rather a key methodological component that enables diverse downstream analyses and validation strategies across cryo-EM and cryo-ET.

\paragraph{Overview of results and contributions.} 
In this work, we introduce a versatile Bayesian framework for orientation estimation with strong statistical guarantees and high flexibility. Section~\ref{sec:problemFormulation} formulates the problem and develops the Bayesian MMSE estimator. We show theoretically that the MMSE estimator coincides with the MLE estimator in high-SNR regimes (Proposition~\ref{thm:mapAndMmseEstimatorsConicideness}), while consistently outperforming it in low-SNR conditions (Figure~\ref{fig:fig1}), a typical scenario in cryo-EM and cryo-ET \cite{bendory2020single}.

Section~\ref{sec:numericalMethods} evaluates the MMSE orientation estimator through simulations. We investigate the effect of different prior distributions and the discretization resolution $L$ of the rotation group $\s{SO}(3)$. Results indicate that non-uniform priors substantially improve estimation accuracy, underscoring the value of incorporating prior knowledge (Figure~\ref{fig:2}). We further show that in high-SNR regimes the estimation error scales as $L^{1/3}$ (Figure~\ref{fig:3}), highlighting discretization as the dominant error source, whereas in low-SNR regimes noise dominates.

In Section~\ref{sec:poseDetrminantionVolumeReconstruction}, we connect the MMSE estimator to 2D image recovery and 3D structure reconstruction, showing that it naturally leads to the expectation-maximization (EM) framework for models without projections (Proposition~\ref{thm:relationBetweenMMSEandEM}). Empirical results confirm that the MMSE estimator consistently outperforms the MLE and MAP estimators in both tasks, providing higher accuracy and greater robustness against the “Einstein from Noise” artifact \cite{henderson2013avoiding, balanov2024einstein} (Figures~\ref{fig:4}–\ref{fig:5}).

Finally, Section~\ref{sec:heterogeneityAnalysis} integrates MMSE-based pose estimation into structural heterogeneity analysis. Following the fixed-pose framework of RECOVAR, we show that MMSE pose estimator consistently improves the recovery of conformational variability compared to the MLE counterpart. Section~\ref{sec:conclutions} concludes with a discussion of further applications of the Bayesian approach to orientation estimation and future research directions.

\paragraph{Main takeaways.}
Figure \ref{fig:fig1} illustrates the critical role of orientation estimation in the cryo-EM and cryo-ET reconstruction processes, showcasing the superior performance of the MMSE orientation estimator, $\hat{g}_{\text{MMSE}}$, compared to the MLE orientation estimator, $\hat{g}_{\text{MLE}}$. Specifically, the curves show that the MMSE estimator consistently produces more accurate estimates than the MLE estimator, with the performance gap widening as the SNR decreases. Moreover, incorporating prior knowledge of \textit{rotation distribution} into our MMSE estimator allows for even better performance (See Figure \ref{fig:2}). This performance gap becomes even more pronounced when the orientation estimators are incorporated into a reconstruction algorithm (see Section~\ref{sec:poseDetrminantionVolumeReconstruction}).

Beyond reconstruction accuracy, we demonstrate that orientation estimation quality has a direct and substantial impact on downstream structural heterogeneity analysis. Understanding structural variability is essential for characterizing the dynamic behavior of macromolecular complexes. Methods such as \textsc{RECOVAR}~\cite{gilles2025cryo}, designed to recover structural heterogeneity, typically assume known particle poses (i.e., fixed-pose methods). In reality, these poses must be inferred, and MLE-derived estimates are commonly used in practice. We show that replacing MLE-derived poses with MMSE estimates in \textsc{RECOVAR} substantially improves the recovery of the latent conformational manifold, bringing results closer to those obtained using ground-truth poses. These findings establish MMSE orientation estimation not only as a tool that enhances reconstruction fidelity, but also as a facilitator that advances state-of-the-art continuous heterogeneity analysis.

The main takeaway of this paper is the recommendation to adopt the Bayesian MMSE orientation estimator for determining the orientation of each observation, in place of the commonly used MLE estimator. The MMSE estimator demonstrates superior performance even under a uniform rotation distribution and offers further improvements when incorporating prior knowledge of the underlying rotation distribution. Importantly, while current software packages do not directly compute the MMSE estimator, they already calculate all the necessary components required for its implementation. Therefore, adopting the MMSE approach can be achieved with minimal additional computational cost. The detailed implementation and code are available at \href{https://github.com/AmnonBa/bayesian-orientation-estimation}{https://github.com/AmnonBa/bayesian-orientation-estimation}.

\section{Problem formulation and the MMSE orientation estimator} \label{sec:problemFormulation}

In this section, we present a particular Bayes estimator for orientation determination under a squared-error loss, commonly known as the \emph{minimum mean-square error} (MMSE) estimator. The name reflects that this estimator minimizes the expected posterior mean-squared error, and it is given by the posterior (conditional) mean rather than a pointwise maximization~\cite{kay1993fundamentals}.

We begin by introducing a flexible mathematical model that encompasses various typical applications involving orientation estimation. Following this, we present a couple of metrics designed to assess the quality of our estimators, providing a robust framework for evaluating performance. Finally, we introduce the class of Bayes estimators, with the MMSE estimator serving as a primary example.

Throughout this paper, we use $g$ to denote both the rotation operator and its corresponding matrix representation. For instance, $g$ can be represented by a three by three rotation matrix in 3D. The intended meaning will be clear from the context, and this slight abuse of notation should not cause confusion.

\subsection{Mathematical model for orientation estimation} \label{sec:mathemticalModel}
We consider a unified framework for modeling measurement processes, encompassing problems such as 2D template matching, orientation estimation in cryo-EM, and subtomogram averaging in cryo-ET. To better focus on the core aspects of our methodology, we omit certain physical effects, such as CTF and in-plane shifts, in this formulation. A complete mathematical model of the cryo-EM imaging process, incorporating these effects, is provided in Appendix \ref{appx:fullCryoEmModel}.

We begin with a continuous-domain formulation. Let \( V (x): \mathbb{R}^n \to \mathbb{R} \) denote a reference structure, where \( n = 2 \) for 2D images and \( n = 3 \) for 3D volumes. For example, when $n=3$, $V$ corresponds to the continuous 3D electron density, as commonly used in cryo-EM or cryo-ET. Let \( \mathsf{G} \subset \mathsf{SO}(n) \) be a compact group of rotations correspondingly, and let \( g \in \mathsf{G} \) be an unknown transformation drawn from a distribution \( \Lambda \) over \( \mathsf{G} \). 

We denote the measurement as a function \( y(x') : \mathbb{R}^m \to \mathbb{R} \), where \( m \leq n \) accounts for possible dimension reduction due to projection (e.g., \( m = 2 \) in cryo-EM). Then, the continuous measurement model is given by
\begin{equation*} 
    y(x') = \Pi\Big( (g \circ V)(x)\Big) + \varepsilon(x'), \qquad x\in\R^n\text{ and } x' \in \mathbb{R}^m,
\end{equation*}
where \( (g \circ V)(x) := V(g^{-1} x) \) is the rotated structure, \( \Pi \) is a known linear operator (such as a tomographic projection or identity), and \( \varepsilon \) denotes additive noise in the measurement domain.

In practice, measurements are only available in discretized form due to finite resolution. In the discrete setting, we assume that the operator \( \Pi \) also incorporates sampling onto a grid of size \( d \), i.e., it includes both projection (if applicable) and discretization. The resulting discrete measurement model becomes
\begin{equation} \label{eq:model}
    y_i = \Pi(g_i \circ V) + \varepsilon_i,
\end{equation}
where \( y_i \in \mathbb{R}^d \), and \( \varepsilon_i \sim \mathcal{N}(0, \Sigma) \) is Gaussian noise with $d\times d$ positive-definite covariance matrix \( \Sigma \). To simplify notation, we omit $\Pi$ when it solely represents the discretization sampling operator and use the same symbol \( y \) to refer to the discretized measurement throughout the paper, unless stated otherwise.

\paragraph{Applications of the model.}

We present three typical examples of this model. In all cases, the goal is to estimate $g$ given the sample $y$, the structure $\barV$, and the covaraince matrix $\Sigma$.

\begin{enumerate}
    \item \textit{2D template matching.} %\amit{what about shifts and CTF in this example?}
    In this case, $\Pi$ is solely the discretization sampling operator, $d=N\times N$ with $N$ the grid size of the 2D image, $g\in\mathsf{G}= \mathsf{SO}(2)$ is a 2D in-plane rotation, and $\barV$ is a given 2D template image. 
    \item \textit{Rotation estimation in cryo-EM}. Here, we consider a special case of \eqref{eq:model} where $\Pi$ comprises both the sampling and tomographic projection operators, $d=N\times N$ with $N$ the grid size of 2D projection images, $g\in\mathsf{G}=\s{SO}(3)$ is a 3D rotation, and $\barV$ is a given 3D volume representing a known reference 3D structure or a well-grounded structure from prior data analyses. %\tamir{again, it's unclear to me if we include CTF or not; we need to be consistent with that} 
    \item  \textit{3D structure alignment in cryo-ET.} In this scenario, we consider a special case of \eqref{eq:model} where $\Pi$ is solely the discretization sampling operator, $d$ corresponds to the total dimension of 3D subtomograms, $g\in\mathsf{G}=\s{SO}(3)$ is a 3D rotation, and $\barV$ is a given 3D volume. Notably, the 3D alignment problem is also a critical step in the computational pipeline of cryo-EM, see, e.g.,~\cite{singer2024alignment,harpaz2023three}.
\end{enumerate}

\subsection{Preliminaries}
Before introducing the specific estimator for the model specified in \eqref{eq:model}, it is instructive to briefly revisit the general Bayesian framework and the concept of the Bayes estimator. This will provide the necessary foundation for understanding the development and analysis of the proposed MMSE estimator, as well as its statistical properties superior to those of the widely used MLE estimator.

\paragraph{Overview of the Bayesian framework and the Bayes estimators.} Suppose that we aim to estimate a true rotation $g\in\mathsf{G}$ drawn from a known prior distribution $\Lambda$. Let $\hat{g}$ be an estimator of $g$ based on a measurement $y$ and let $\mathcal{L} \p{g, \hat{g}}$ be a loss function.
%(for example, the chordal or geodesic distances)
The \textit{Bayes risk} of $\hat{g}$ is defined as $\E_\Lambda[\mathcal{L} \p{g, \hat{g}}]$, where the expectation is taken over the data generation process of $y$ given $g$ and the prior distribution $\Lambda$ of $g$. The \textit{Bayes estimator} with respect to the loss $\mathcal{L}$ \cite[Chapter 4, Theorem 1.1]{lehmann2006theory} is defined as the estimator that minimizes the Bayes risk among all possible estimators, i.e.,
\begin{align}
    \hat{g}_{\s{Bayes}} = \underset{\hat{g}} {\argmin} \ {\mathbb{E}_{\Lambda}\pp{\mathcal{L} \p{g, \hat{g}}}}.
\end{align}
Equivalently, it is the estimator that minimizes the posterior expected loss $\E_{\Lambda}[\mathcal{L} \p{g, \hat{g}}|y]$, where the expectation is taken over the posterior distribution of $g$ given the measurement $y$, and any other known parameter such as volume $\barV$, with the prior $\Lambda$. For the case of $G = \s{SO}(3)$, it is given explicitly by
\begin{align}
    \nonumber \hat{g}_{\s{Bayes}} & = \underset{\hat{g}\in \s{SO}(3)} {\argmin} \ \mathbb{E}_{\Lambda} \pp{\mathcal{L} \p{g, \hat{g}}| y, \barV} 
    \\ & = \underset{\hat{g}\in \s{SO}(3)} {\argmin} \int_{\s{SO(3)}} \mathcal{L} \p{g, \hat{g}} \P_\Lambda(g|y, \barV)\der g , \label{eqn:bayesEstDef}
\end{align}
where by Bayes' law, we have, 
\begin{align}
\P_\Lambda(g|y, \barV)\der g = \frac{\P(y|g, \barV) \der\Lambda(g) }{\int_{\mathsf{G}}\P(y|g, \barV)\der\Lambda(g)}. \label{eqn:bayesRule}
\end{align}

\paragraph{The posterior distribution of the rotation $g$ given an observation $y$ under model \eqref{eq:model}.}
To introduce the MMSE estimator corresponding to the model \eqref{eq:model}, we first compute the posterior distribution of $g$ given $y$ and all the additional parameters $\barV$, and $\Sigma$.
We obtain the conditional likelihood density
\begin{align} \label{eq:liklihood_density}
    \P(y|g,\barV)= \frac{1}{(2\pi)^{d/2} \mathrm{det}({\Sigma})^{1/2}}\exp\Big(-\frac{1}{2}(y-\Pi (g\circ \barV))^\sT \Sigma^{-1} (y-\Pi (g\circ \barV))\Big).
\end{align}
Note that $g$ follows the underlying prior distribution $\Lambda$.
Applying Bayes' law, the posterior distribution of $g$ given $y$ is 
\begin{align}\label{eq:posterior_pose}
    \P_{\Lambda}(g|y,\barV) \der g = \frac{\exp\Big(-\frac{1}{2}(y-\Pi (g\circ \barV))^\sT \Sigma^{-1} (y-\Pi (g\circ \barV))\Big)\der \Lambda(g)}{\int_{\mathsf{G}} \exp\Big(-\frac{1}{2}(y-\Pi (g\circ \barV))^\sT \Sigma^{-1} (y-\Pi (g\circ \barV))\Big) \der \Lambda(g)}.
\end{align}

\paragraph{Metrics over SO(3).}
As the Bayes estimator is closely related to the given loss function $\mathcal{L}$, we present two candidate metrics on the 3D special orthogonal group $\s{SO}(3)$. Here we represent any rotation $g \in \s{SO}(3)$ in its natural three by three matrix representation.

\begin{enumerate}
    \item \textit{Chordal distance.} For any two rotations $g_1,g_2\in\s{SO}(3)$, the chordal distance is defined as 
    \begin{align}\label{eq:chor_dist}
    d_F (g_1,g_2) \triangleq \|g_1-g_2\|_\F = \sqrt{\tr \Big({\p{g_1-g_2}^\sT\p{g_1-g_2}}\Big)},
    \end{align}
    where $\norm{\cdot}_\F$ represents the matrix Frobenius norm, and $\tr(\cdot)$ is the trace of a matrix. This metric is easy to compute and analyze; however, it does not take into account the group structure of rotations.
    
    \item \textit{Geodesic distance.} For any two rotations $g_1,g_2\in\s{SO}(3)$, the geodesic distance is defined as 
    \begin{align}\label{eq:geo_dist}
    d_G (g_1,g_2) \triangleq \arccos \p{\frac{\tr(g_2g_1^{-1})-1}{2}},
    \end{align}
    where $\tr(\cdot)$ is the trace of a matrix. This metric reflects the shortest path between $g_1$ and $g_2$ in the 3D rotation manifold.
\end{enumerate}

We note that the chordal and geodesic distances are closely related. In particular, if $\theta \triangleq d_G(g_1,g_2)$ denotes the geodesic rotation angle between $g_1$ and $g_2$~\eqref{eq:geo_dist}, then
\begin{align}
    d_F(g_1,g_2) = \|g_1-g_2\|_F = 2\sqrt{2}\,\sin(\theta/2) = 2\sqrt{2}\,\sin\!\big(d_G(g_1,g_2)/2\big).
\end{align}
Consequently, $d_F$ is a strictly increasing function of $d_G$ and, for small angular errors, the two are locally equivalent in the sense that
$d_F(g_1,g_2)\approx \sqrt{2}\,d_G(g_1,g_2)$ (see, e.g., \cite{hartley2013rotation}).
More generally, our framework is flexible and can accommodate other loss functions over any group $\mathsf{G}$. For a comprehensive discussion of metrics on $\mathsf{SO}(3)$ and their relationships, we refer the reader to \cite{hartley2013rotation,huynh2009metrics}. Similar considerations apply to $\mathsf{SO}(2)$; we omit a detailed discussion for brevity.

\subsection{The MLE, MAP and MMSE orientation estimators} \label{sec:mmseMapAndMLest}

In the following presentation, we focus on independent and identically distributed Gaussian noise with $\Sigma = \sigma^2 I_{d \times d}$. These noise statistics are commonly employed in modern software tools~\cite{kimanius2021new}. While the proposed framework can incorporate more advanced noise models, we restrict attention to this case for clarity of presentation and to simplify the theoretical analysis.

\paragraph{The MLE estimator.}
Recalling the maximum cross-correlation method we mentioned earlier, we now connect it with the mathematical model in \eqref{eq:model} and introduce the corresponding MLE estimator. 
The rotation $g$ that minimizes the distance between the corresponding projected rotated volume $\Pi(g \circ \barV)$ and the observation $y$ is exactly the MLE estimator defined as
\begin{align}
    \nonumber \hat{g}_{\s{MLE}}:=& \argmax_{g\in \s{SO}(3)} \P(y|g,\barV) \\ 
    \nonumber  =& \argmax_{g\in \s{SO}(3)}   \exp\Big(-\|y-\Pi( g \circ \barV)\|^2/2\sigma^2\Big)\\ 
     =&\argmin_{g\in \s{SO}(3)}   \|y-\Pi( g \circ \barV)\|^2\label{eq:mlEstimator},
\end{align}
where the first equality follows from \eqref{eq:liklihood_density}, and the second equality follows from the assumption $\Sigma=\sigma^2 I_{d\times d}$. In other words, $\hat{g}_{\s{MLE}}$ maximizes the conditional density \eqref{eq:liklihood_density} over all possible rotations in $\s{SO}(3)$. In the absence of the tomographic projection (e.g., in cryo-ET), the MLE estimator further simplifies to 
\begin{align*}
    \hat{g}_{\s{MLE}}= \argmax_{g\in \s{SO}(3)} y^{\sT} ( g \circ \barV),
\end{align*}
which corresponds to the rotation $g$ such that the rotated structure maximizes the correlation with~$y$. 
This estimator is also frequently used in single-particle cryo-EM, under the assumption that the norm of $\Pi( g \circ \barV)$ is approximately constant for all $g$. 
This estimator can be approximated by performing a search over a pre-defined grid of 3D rotations, selecting the rotation that minimizes the distance between the measurement $y$ and the projected rotated volume $\Pi( g \circ \barV)$. This approach forms the basis of standard practices in single-particle cryo-EM and cryo-ET.

\paragraph{The MAP estimator.} The MAP estimator extends the MLE estimator, by incorporating prior knowledge on the distribution of the rotation in the special group $\s{SO}(3)$, $\der\Lambda(g)$. Formally, the MAP estimator is defined as
\begin{align}
    \nonumber \hat{g}_{\s{MAP}} & = \argmax_{g\in \s{SO}(3)} \P_{\Lambda}(g|y,\barV) 
    \\ \nonumber & = \argmax_{g\in \s{SO}(3)} \frac{\P(y|g, \barV) \der\Lambda(g) }{\int_{\mathsf{G}}\P(y|g, \barV)\der\Lambda(g)}
    \\ \nonumber & = \argmax_{g\in \s{SO}(3)} \P(y|g,\barV) \der\Lambda(g)
    \\ \nonumber & = \argmax_{g\in \s{SO}(3)}   \exp\Big(-\|y-\Pi( g \circ \barV)\|^2/2\sigma^2\Big) \der\Lambda(g)
    \\ & = \argmin_{g\in \s{SO}(3)}   \|y-\Pi( g \circ \barV)\|^2 /2{\sigma}^2 - \log  \p{\der\Lambda(g)}, \label{eq:mapEstimator}
\end{align}
where the second equality follows from Bayes' law \eqref{eqn:bayesRule}, and the third equality from independence of the denominator on $g$. In other words, $\hat{g}_{\s{MAP}}$ maximizes the posterior density \eqref{eq:posterior_pose} over all possible rotations in $\s{SO}(3)$. It can be seen that the MAP estimator $\hat{g}_{\s{MAP}}$ \eqref{eq:mapEstimator} coincide with the MLE estimator $\hat{g}_{\s{MLE}}$ \eqref{eq:mlEstimator} when the prior distribution $\der\Lambda(g)$ is uniform over $\s{SO}(3)$.

\paragraph{The MMSE estimator.}
For any rotation distribution $\Lambda$ and loss function $\mathcal{L}(\cdot,\cdot)$ over $\s{SO}(3)$, the MLE and MAP estimators can be further improved by the corresponding Bayes estimator which minimizes the posterior expected loss $\E_{\Lambda}[\mathcal{L}(g,\cdot)\mid y,\barV]$. In particular, following the definition \eqref{eqn:bayesEstDef} of the Bayes estimator, and for the chordal distance $d_F(\cdot,\cdot)$ \eqref{eq:chor_dist}, with its corresponding squared loss $\mathcal{L}_F(\cdot,\cdot)=d_F^2(\cdot,\cdot)$, the Bayes estimator takes the form
\begin{align}
    \hat{g}_{\s{MMSE}}
    := \underset{\hat{g} \in \s{SO}(3)} {\argmin} \ \mathbb{E}_{\Lambda} \!\left[d_F^2 \p{g, \hat{g}} \mid y, \barV \right], \label{eqn:mmseEstimator}
\end{align}
where the expectation is taken over the posterior density \eqref{eq:posterior_pose}. Here we denote the estimator as $\hat{g}_{\s{MMSE}}$, since minimizing the squared chordal distance is equivalent to minimizing the entrywise mean-square error between the two $3\times 3$ rotation matrices. Alternative distance measures, such as the geodesic distance on $\s{SO}(3)$ discussed earlier, could certainly be considered. In this work, however, we focus on the chordal distance, primarily because it admits a particularly convenient derivation of the associated Bayes estimator, as shown below, thereby offering both computational efficiency and interpretability. Importantly, these Bayes estimators are generally distinct from both the MLE and MAP estimators—even under a uniform prior—since it minimizes the posterior expected loss rather than selecting the most probable rotation.

To compute $\hat{g}_{\s{MMSE}}$, we begin by noting that it is defined over the manifold $\s{SO}(3)$. In general, directly solving the optimization problem on a non-Euclidean manifold is challenging. To address this, we first relax the optimization domain to the ambient Euclidean space $\mathbb{R}^{3\times 3}$, which leads to the following intermediate estimator:
\begin{align}
\hat{g}_{\s{relax}}
&:= \underset{\hat{g} \in \mathbb{R}^{3 \times 3}} {\argmin} \ \mathbb{E}_{\Lambda}\!\left[d_F^2 \p{g, \hat{g}} \,\middle|\, y, \barV\right] \notag\\
&= \E_{\Lambda}[g \mid y,\barV] \notag\\
&=\frac{\int_{\s{SO}(3)} g \exp\!\Big(-\tfrac{1}{2}(y-\Pi (g\circ \barV))^\sT \Sigma^{-1} (y-\Pi (g\circ \barV))\Big)\,\der \Lambda(g)}{\int_{\s{SO}(3)} \exp\!\Big(-\tfrac{1}{2}(y-\Pi (g\circ \barV))^\sT \Sigma^{-1} (y-\Pi (g \circ \barV))\Big) \,\der \Lambda(g)} \notag\\
&=\frac{\int_{\s{SO}(3)} g \exp\!\Big(-\tfrac{1}{2\sigma^2}\big\|y-\Pi (g\circ \barV)\big\|_2^2\Big)\,\der \Lambda(g)}{\int_{\s{SO}(3)} \exp\!\Big(-\tfrac{1}{2\sigma^2}\big\|y-\Pi (g\circ \barV)\big\|_2^2\Big) \,\der \Lambda(g)}.\label{eqn:relaxEstimator}
\end{align}
Here, since we naturally represent $g \in \s{SO}(3)$ as a $3\times 3$ rotation matrix, the expectation (or integral) can be interpreted directly in this matrix representation, yielding a $3\times 3$ matrix as the posterior mean matrix\footnote{More generally, if one works with different representations of $\s{SO}(3)$, or interprets the expectation as acting on operators (as in Proposition~\ref{thm:relationBetweenMMSEandEM}), the resulting estimator may take different forms.}. However, $\hat g_{\s{relax}}$ does not, in general, lie in $\s{SO}(3)$ and hence may not be a valid rotation operator. To obtain a feasible solution on the manifold, we subsequently apply the \textit{orthogonal Procrustes} procedure to project $\hat g_{\s{relax}}$ back onto $\s{SO}(3)$, as discussed next.

Given a matrix $A$, to obtain a valid rotation matrix we seek $\Omega \in \s{SO}(3)$ that is closest to $A$ in the Frobenius norm:
\begin{align}
    \underset{\Omega}{\text{minimize}} \quad \|\Omega - A\|_F \quad  
    \text{subject to} \quad \Omega \in \s{SO}(3).\label{eqn:orthogonalProcrustes}
\end{align}
This optimization problem is known as the \textit{Orthogonal Procrustes} problem in the linear algebra literature. 
It can be solved efficiently using the singular value decomposition (SVD) of $A$~\cite{gower2004procrustes}.     Concretely, given a matrix $A\in\mathbb{R}^{3\times 3}$ (in our case $A=\hat{g}_{\mathrm{relax}}$), compute its singular value decomposition
\begin{align}
    A = U \Sigma V^{\top}, \qquad U,V\in O(3),\ \Sigma\succeq 0.
\end{align}
The closest rotation matrix in Frobenius norm is then obtained by
\begin{align}
    \Omega = U\,\mathrm{diag}\!\bigl(1,\,1,\,\det(UV^{\top})\bigr)\,V^{\top}\ \in \mathsf{SO}(3).
\end{align}
Equivalently, if $\det(UV^{\top})=-1$, one flips the sign of the last column of $U$ (or $V$) before forming $UV^{\top}$, ensuring $\det(\Omega)=1$ and yielding the Procrustes solution. Thus, applying the Orthogonal Procrustes procedure to $\hat g_{\s{relax}}$ projects it back onto $\s{SO}(3)$, yielding a valid rotation matrix. 
The following proposition establishes that this valid rotation matrix coincides with the MMSE estimator $\hat{g}_{\s{MMSE}}$, 
thereby completing the procedure for obtaining the MMSE estimate. 
A detailed proof is provided in Appendix \ref{appx:proofOrthogonalProcrustes}.

\begin{proposition}
\label{thm:orthogonalProcrustes}
The Bayes estimator of the loss function defined by the chordal distance $\hat{g}_{\s{MMSE}}$ \eqref{eqn:mmseEstimator} is equal to the Orthogonal Procrustes solution \eqref{eqn:orthogonalProcrustes} applied on the intermediate estimator $\hat{g}_{\s{relax}}$, i.e.,
\begin{align}
    \hat{g}_{\s{MMSE}} = {\argmin}_{\Omega \in \s{SO}(3)} \ \norm{\hat{g}_{\s{relax}} - \Omega}_F. \label{eqn:propOrthProc}
\end{align}
\end{proposition}

\paragraph{The MMSE, MAP and MLE estimators in the high SNR regime.} We conclude the section by showing the next proposition. It shows that the MMSE and MAP estimators converge in the high SNR regime (i.e., $\sigma \to 0$), as demonstrated empirically in Figures \ref{fig:3}, \ref{fig:4}, and \ref{fig:5}. This implies that the MMSE estimator's superior statistical properties are most advantageous in low SNR conditions, which are common in structural biology applications like cryo-EM and cryo-ET \cite{bendory2020single}. In these low SNR environments, the MMSE estimator consistently outperforms its MAP counterpart.  

\begin{proposition}
\label{thm:mapAndMmseEstimatorsConicideness}
Let $\Lambda$ be any distribution over $\s{SO}(3)$ with a strictly positive density, bounded below by some constant $c>0$. Let $\hat{g}_{\s{MLE}}$, $\hat{g}_{\s{MAP}}$ be the MLE and MAP rotation estimators as defined in \eqref{eq:mlEstimator} and \eqref{eq:mapEstimator}, respectively, and assume they are unique. Let $\hat{g}_{\s{MMSE}}$ be the MMSE estimator as defined in \eqref{eqn:mmseEstimator}. Then, we have,
     \begin{align}
        \lim_{\sigma \to 0} \hat{g}_{\s{MMSE}}=\lim_{\sigma \to 0} \hat{g}_{\s{MAP}} = \hat{g}_{\s{MLE}}.
     \end{align}
\end{proposition}
The proof of the proposition is provided in Appendix \ref{apx:mapAndMmseEstimatorsConicideness}. The proof relies on the existing result,~\cite[Theorem 5.10]{robert1999monte}, and is valid not only to rotations but also to other group operators, such as translations.

\section{Numerical methods for MMSE orientation estimation} \label{sec:numericalMethods}
This section compares the numerical performance of the MMSE and MLE estimators. We also introduce various types of prior distributions for $\s{SO}(3)$ beyond the uniform distribution and demonstrate how incorporating this prior knowledge can significantly improve the performance of the MMSE rotation estimator. Furthermore, we study the influence of the number of sampling points of the group $\s{SO}(3)$ of 3D rotations (namely, the number of candidate rotations), denoted as $L$, on the quality of rotation estimation. For better illustration, we consider the simplified setting of model \eqref{eq:model}, where $\Sigma = \sigma^2 I_{d\times d}$, although our method applies to the general covariance setting as well.

\paragraph{Numerical procedure and sampling of $\s{SO}(3)$.}
Since the posterior distribution of $g \in \s{SO}(3)$ in~\eqref{eq:posterior_pose} is continuous, a numerical discretization of the rotation group is required. 
In this work, the expectation is approximated using a quadrature rule over $\s{SO}(3)$. 
Let $\{(g^{(\ell)}, w_\ell)\}_{\ell=0}^{L-1}$ denote a numerical quadrature on $\s{SO}(3)$,
constructed as a product of a spherical quadrature on $\s{S}^2$ and a circular rule over the Euler angles~\cite{graf2009sampling,
graf2012unified}.
The implementation employed here follows the publicly available MATLAB code described in~\cite{hoskins2024subspace}.

We first consider the uniform case, where $\Lambda$ is the Haar distribution on $\s{SO}(3)$. For each quadrature node, define $x_{\ell} := \Pi\!\big(g^{(\ell)} \circ \barV\big)$, for  $0 \leq \ell \leq L-1$, where $\{g^{(\ell)}\}_{\ell=0}^{L-1}$ are the quadrature nodes and $\{w_\ell\}_{\ell=0}^{L-1}$ are their associated weights.
The MMSE estimator can then be expressed as the weighted average
\begin{align}
    \hat{g}_{\s{MMSE}}
    = \E[g \mid y, \barV]
    &\approx  
    \frac{\displaystyle \sum_{\ell=0}^{L-1} g^{(\ell)}\, w_\ell 
    \exp\!\Big(-\tfrac{\|y - x_\ell\|^2}{2\sigma^2}\Big)}
         {\displaystyle \sum_{\ell=0}^{L-1} w_\ell 
    \exp\!\Big(-\tfrac{\|y - x_\ell\|^2}{2\sigma^2}\Big)}
    = \sum_{\ell=0}^{L-1} g^{(\ell)}\, p^{(\ell)},
    \label{eqn:g_MMSE_numerical}
\end{align}
where the posterior quadrature weights are defined as
\begin{align}
    p^{(\ell)} 
    \;\triangleq\;
    \frac{w_\ell 
        \exp\!\Big(-\tfrac{\|y - x_\ell\|^2}{2\sigma^2}\Big)}
        {\displaystyle \sum_{r=0}^{L-1} 
        w_r \exp\!\Big(-\tfrac{\|y - x_r\|^2}{2\sigma^2}\Big)},
    \qquad 0 \leq \ell \leq L-1.
    \label{eq:posterior_pose_numerical}
\end{align}

For a general non-uniform prior $\Lambda(g)$, the same quadrature can be applied with modified posterior weights:
\begin{align}
    p^{(\ell)} \;\propto\;
    w_\ell\, \Lambda\!\big(g^{(\ell)}\big)\,
    \exp\!\Big(-\tfrac{\|y - x_\ell\|^2}{2\sigma^2}\Big),
\end{align}
followed by normalization to ensure $\sum_{\ell=0}^{L-1} p^{(\ell)} = 1$.

Similarly, the MLE estimator in \eqref{eq:mlEstimator} relies on the discretization of $\s{SO}(3)$, which can be approximated through grid search using the uniform rotation samples or pre-defined grid $\ppp{g^{(\ell)}}_{\ell=0}^{L-1}$ as 
\begin{align}
    \hat{g}_{\s{MLE}} =& \underset{0 \leq \ell \leq L-1} {\argmin} \norm{y-x_\ell}^2. \label{eqn:g_MAP_discrete}
\end{align}
Both estimators require evaluating $\|y - x_\ell\|$ for all candidate rotations, which naively costs $O(d)$ per candidate and thus $O(Ld)$ overall. This estimate, however, overlooks the structure of $\s{SO}(3)$. Any 3D rotation can be decomposed into a \emph{viewing direction} (two degrees of freedom, discretized with about $L^{2/3}$ samples) and an \emph{in-plane rotation} about that direction (one degree of freedom, discretized with $L^{1/3}$ samples). By exploiting FFT-based methods, all $L^{1/3}$ in-plane rotations can be evaluated simultaneously for each viewing direction. Consequently, the effective complexity becomes $O(L^{2/3} \, d \, \log d)$, which improves over the naive bound by a factor of $L^{1/3}$~\cite{kostelec2008ffts,kileel2024fast}. This reduction is substantial in practice, since $L$ often reaches tens or even hundreds of thousands in high-resolution cryo-EM.

\begin{algorithm}[t!]
  \caption{\texttt{Numerical solver for MMSE rotation estimator} \label{alg:mmseEstimator}}
\textbf{Input:} Observation $y$ following \eqref{eq:model}, and a set of candidate rotations $\ppp{g^{(\ell)}}_{\ell=0}^{L-1}$ of $\s{SO}(3)$.\\
\textbf{Output:} MMSE estimator of the rotation, $\hat{g}_{\s{MMSE}}$.
\begin{enumerate}
    \item Compute:
        \begin{align}
            \hat{g}_{\s{relax}} = \frac{\sum_{\ell=0}^{L-1} g^{(\ell)} \exp\Big(-\norm{y-\Pi \p{{g^{(\ell)}} \circ\barV}}^2/2\sigma^2\Big)}{\sum_{\ell=0}^{L-1} \exp\Big(-\norm{y- \Pi \p{{g^{(\ell)}} \circ \barV}}^2/2\sigma^2\Big)}.\label{eqn:g_MMSE_numerical_al2g}
        \end{align}
    \item Perform the Orthogonal Procrustes algorithm \eqref{eqn:orthogonalProcrustes} on $\hat{g}_{\s{relax}}$ in \eqref{eqn:g_MMSE_numerical_al2g} to get the proper rotation estimator $\hat{g}_{\s{MMSE}}$.
\end{enumerate}
\end{algorithm}

\paragraph{The impact of non-uniform distributions of rotations.} One key advantage of the Bayesian framework is its flexibility in incorporating different prior distributions for the rotations. In practical cryo-EM applications, the distribution of particle orientations is often non-uniform due to preferred particle orientations or sample preparation artifacts~\cite{tan2017addressing,lyumkis2019challenges}. If the rotation distribution can be well estimated in an early stage, the information of the rotation distribution can then be integrated into the Bayes estimator to improve rotation estimation. 

To illustrate rotation estimation under different prior distributions on $\s{SO}(3)$, we replace the uniform distribution on $\s{SO}(3)$ with an isotropic Gaussian (IG) distribution on $\s{SO}(3)$, denoted by $g \sim \mathcal{IG}_{\s{SO}(3)}(\eta)$, parameterized by a scalar variance $\eta^2$. This distribution is frequently used in machine learning probabilistic models on $\s{SO}(3)$ \cite{corso2022diffdock, leach2022denoising, jagvaral2024unified}. It is worth noting that the IG distribution serves as a typical example, and similar phenomena as observed here extend to other non-uniform distributions on $\s{SO}(3)$ as well.

The IG distribution $\mathcal{I}\mathcal{G}_{\s{SO}(3)}(\eta)$ can be represented in an axis-angle form, with uniformly sampled axes of rotation and a rotation angle $\omega \in [0, \pi]$. 
The scalar variance $\eta^2$ controls the distribution of the rotation angle $\omega$: as $\eta\to \infty$, the $\s{SO}(3)$ distribution approaches uniformity, whereas as $\eta \to 0$, $\omega$ becomes increasingly concentrated around 0, i.e., the rotation angle around the rotation axis is small. We apply the inverse sampling method to obtain i.i.d. samples from the IG distribution. Further details on this distribution are provided in Appendix \ref{apx:isotropicGaussianDef}.

Figure \ref{fig:2} illustrates how incorporating a prior distribution over $\s{SO}(3)$ rotations, governed by the variance parameter $\eta$ of an isotropic Gaussian distribution affects, the accuracy of the MMSE rotation estimator. In all cases, the true underlying rotation distribution is modeled as an isotropic Gaussian distribution $\mathcal{IG}_{\s{SO}(3)}(\eta=0.1)$ over $\s{SO}(3)$. The MLE estimator was computed according to \eqref{eqn:g_MAP_discrete}. For the MMSE estimators, the estimation process used different prior distributions with variance parameters $\eta=0.7,0.5$, and $0.1$, respectively. Specifically, the candidate rotations $g^{(\ell)}$ were generated according to these different priors (see \eqref{eqn:g_MMSE_numerical}), highlighting the impact of prior mismatch on estimation accuracy. As the variance decreases (indicating a more concentrated and less uniform distribution closer to the true underlying distribution), the performance of the MMSE estimator improves, particularly at lower SNR conditions. These findings highlight the value of incorporating prior knowledge in rotation estimation, demonstrating its potential to enhance accuracy substantially. In stark contrast, the MLE estimator remains entirely unchanged regardless of the true underlying rotation distribution. As a result, its performance remains suboptimal, particularly in scenarios where the true distribution deviates significantly from uniformity.

We note that in experimental cryo-EM/ET datasets the viewing-angle distribution is typically unknown. Estimating this distribution, and understanding how \emph{prior misspecification} impacts downstream reconstruction and heterogeneity analysis remains an important open problem. We discuss possible approaches for estimating the rotational prior in Appendix~\ref{sec:prior-dist-est}. Recent work has begun to quantify misspecification effects under non-uniform group actions; see, e.g., \cite{xu2025misspecified}.  In this light, Figure~\ref{fig:2} can be viewed as a controlled \emph{prior-mismatch} study: the data are generated from a concentrated isotropic-Gaussian prior, whereas the MMSE (and MAP) estimators are evaluated using an assumed prior with varying concentration. We observe a graceful degradation under mismatch, and a clear improvement as the assumed prior approaches the true distribution. Importantly, even under the common assumption of a uniform prior, the MMSE estimator improves over MLE in the low-SNR regime, while more accurate prior information yields additional gains when strong preferred orientations are present.

\begin{figure}[t!]
    \centering
    \includegraphics[width=0.8\linewidth]{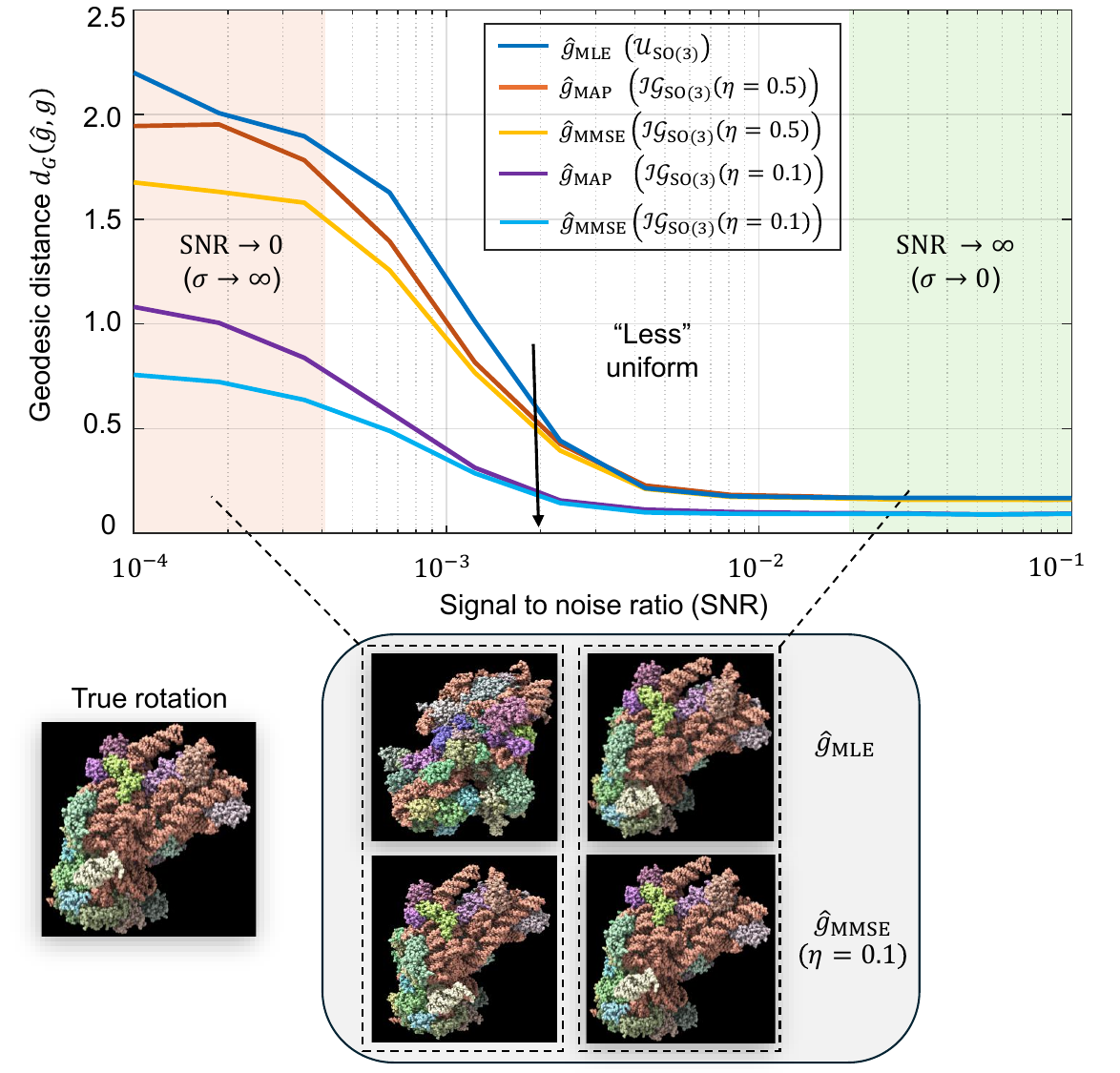}
    \caption{\textbf{Impact of incorporating the prior rotation distribution on estimation accuracy.}
    Simulations are performed for the cryo-ET model~\eqref{eq:cryoET}, excluding the projection step.
    The true rotation distribution is modeled as an isotropic Gaussian $\mathcal{IG}_{\mathsf{SO}(3)}(\eta=0.1)$. Estimation performance is measured using the geodesic distance defined in~\eqref{eq:geo_dist}. Here, $g$ denotes the true rotation, $\hat{g}_{\mathrm{MLE}}$ is the maximum-likelihood estimator from~\eqref{eqn:g_MAP_discrete}, $\hat{g}_{\mathrm{MAP}}$ is the maximum a posteriori estimator from~\eqref{eq:mapEstimator}, and $\hat{g}_{\mathrm{MMSE}}$ denotes the Bayesian minimum mean square error estimator from~\eqref{eqn:g_MMSE_numerical}.    
    The MMSE estimators are computed assuming isotropic Gaussian priors on $\mathsf{SO}(3)$ with different concentration parameters $\eta \in \{0.5, 0.1\}$ (see Appendix~\ref{apx:isotropicGaussianDef}). As $\eta$ decreases, the prior becomes more concentrated and closer to the true underlying distribution, leading to improved accuracy of both the MAP and MMSE estimators. Each data point in the plot is averaged over 3000 Monte Carlo trials using a rotation grid of size $L=2976$.    
    \textbf{Bottom $2\times 2$ panel.} The four images compare denoised 3D volumes obtained using different rotation estimators at two representative noise regimes marked on the curve plot. Rows correspond to the estimator: top row uses $\hat{g}_{\mathrm{MLE}}$, bottom row uses $\hat{g}_{\mathrm{MMSE}}$ with the true prior $\mathcal{IG}_{\mathsf{SO}(3)}(\eta=0.1)$. Columns correspond to SNR: left column is a low-SNR example, and right column is a high-SNR example (where the true rotation is correctly classified).}
    \label{fig:2}
\end{figure}

\begin{figure}[t!]
    \centering
    \includegraphics[width=0.8\linewidth]{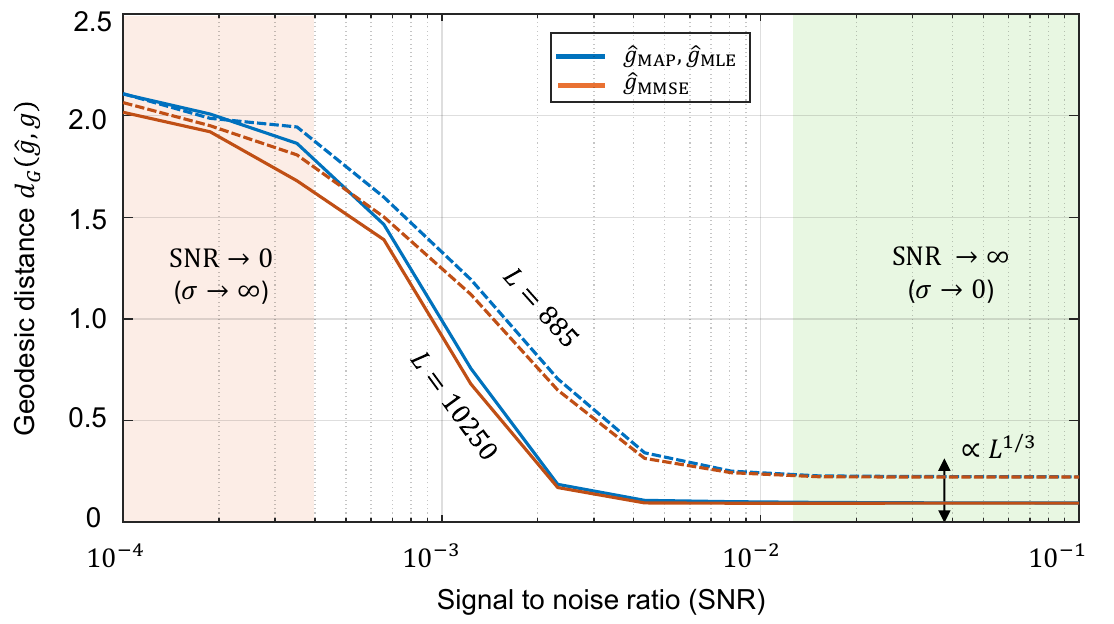}
        \caption{\textbf{Impact of the sampling grid size of $\s{SO}(3)$ ($L$) and the signal-to-noise ratio (SNR) on rotation estimation accuracy.}         
        This figure shows the accuracy of rotation estimation under varying sampling grid sizes $L$ of the rotation group $\s{SO}(3)$ and different SNR levels of the observed data $y$ in the model \eqref{eq:model}. Simulations are performed for the cryo-ET model~\eqref{eq:cryoET}, excluding the projection step. The metric used for comparison is the geodesic distance, as defined in \eqref{eq:geo_dist}. Here, $g$ denotes the true rotation, $\hat{g}_{\mathrm{MLE}}$ represents the MLE estimator from \eqref{eqn:g_MAP_discrete}, and $\hat{g}_{\mathrm{MMSE}}$ denotes the Bayesian MMSE estimator from \eqref{eqn:g_MMSE_numerical}.  
        In the high SNR regime ($\sigma \to 0$), the MLE and MMSE estimators converge, and the geodesic distance scales empirically as $\propto L^{1/3}$. This scaling reflects the three-parameter nature of $\s{SO}(3)$ rotations, where the resolution of the sampling grid improves as $L$ increases. The results shown are based on Monte Carlo simulations with 3000 trials per data point.}
        \label{fig:3}
\end{figure}

\paragraph{The impact of sampling grid size $L$ and SNR on estimation accuracy.} To study these effects, we restrict ourselves to the \textit{uniform} rotation distribution setting. Figure~\ref{fig:3} illustrates the impact of the sampling grid size of $\s{SO}(3)$ together with different levels of SNR, on the geodesic distance between the MLE and MMSE rotation estimators relative to the true rotations. Several observations can be made. First, at high SNR, the MLE and MMSE estimators nearly coincide, consistent with Proposition~\ref{thm:mapAndMmseEstimatorsConicideness}. In this regime, the geodesic error decreases with the grid resolution, scaling empirically as $L^{1/3}$. This behavior arises because a discretization of $\s{SO}(3)$ involves three angular parameters, so the resolution in each parameter direction grows like $L^{1/3}$. Second, as the SNR decreases, noise dominates and the advantage of refining the grid diminishes: both estimators approach similar performance, and the dependence on $L$ becomes weaker. In the extreme limit $\sigma \to \infty$, the mean geodesic distance of the MLE and MMSE estimators is indistinguishable and shows no dependence on the grid size $L$.

\paragraph{Sampling resolution and spectral information.}
Another factor that influences the discrepancy between the MAP and MMSE estimators is the effective spectral information available at a given sampling resolution.
When comparing MAP and MMSE estimates across different image sizes, it is important to distinguish increasing the number of pixels from increasing usable spectral content. If the underlying signal is effectively bandlimited and the image is already sampled at (or above) the Nyquist rate for that bandlimit, then increasing the sampling resolution mainly amounts to oversampling and should not materially change the likelihood or the posterior over $\mathsf{SO}(3)$; consequently, the MAP--MMSE estimators are expected to remain essentially unchanged. In contrast, when higher sampling resolution is accompanied by additional usable high-frequency signal (i.e., non-negligible per-frequency SNR at higher bands), the posterior typically becomes more concentrated, and we expect the MAP and MMSE estimates to approach each other in geodesic distance (at a fixed SNR level).

\section{Bayesian orientation estimation as part of the volume reconstruction problem} \label{sec:poseDetrminantionVolumeReconstruction}

Thus far, we have introduced the MLE, MAP and MMSE orientation estimators for estimating the rotation between a single noisy observation $y$ and a reference volume $\barV$, as defined in \eqref{eq:model}. An intriguing question that arises is how these estimators can be utilized and influence performance within a 3D volume reconstruction process, which constitutes the main computational challenge in cryo-EM and cryo-ET.

Structure reconstruction typically follows two primary approaches: \textit{hard-assignment} or \textit{soft-assignment} methods, which are generally implemented through iterative refinement. In the hard-assignment approach, each observation is assigned a single orientation based on the highest correlation, and the 3D structure is then reconstructed given the rotations. In contrast, the soft-assignment method assigns probabilities across all possible orientations for each observation, enabling the 3D structure to be recovered as a weighted average of the observations, with weights determined by these probabilities. The iterative application of the soft-assignment procedure aligns with the EM algorithm~\cite{dempster1977maximum,sigworth2010introduction}, which serves as the core computational method in modern cryo-EM~\cite{scheres2012relion,punjani2017cryosparc}.

In the following, we demonstrate that incorporating the MMSE orientation estimator into the volume reconstruction process in the cryo-ET model (without projections) resembles the expectation-maximization (EM) algorithm, as it accounts for the full distribution of possible outcomes. In contrast, substituting the MLE estimator into the algorithm operates more like a hard-assignment reconstruction method, focusing exclusively on the most likely outcome. This distinction underscores the broader applicability and flexibility of the Bayes estimator in capturing uncertainty and delivering more accurate estimates for structure reconstruction.

\subsection{Connection to the EM algorithm in volume reconstruction without projection}\label{sec:connection} 

Unlike the previous orientation estimation model \eqref{eq:model}, we consider the following simplified model for volume reconstruction (see also \cite{singer2018mathematics, bandeira2023estimation, fan2024maximum}). We observe $M$ i.i.d samples taking the form 
\begin{align}\label{eq:model_reconstruction}
    y_i = g_i \circ V + \varepsilon_i, \quad\quad i=0,\ldots,M-1,
\end{align}
where $V$ is 3D volume structure of interest, $\{g_i\}_{i=0}^{M-1} \in \s{SO}(3)$ are unknown latent variables following i.i.d. \textit{uniform} distribution, satisfying $g_i \circ V(x) \equiv V(g_i^{-1} x)$, and $\{\varepsilon_i\}_{i=0}^{M-1}$ are i.i.d isotropic Gaussian noise with variance $\sigma^2$. For the case where $V$ represents a 2D image, $\{g_i\}_{i=0}^{M-1} \in \s{SO}(2)$ corresponds to in-plane rotations, and the model is used for image recovery (see more in Section \ref{sec:empirical_reconstruction})~\cite{ma2019heterogeneous}. The goal is to recover $V$ from the observations $\{y_i\}_{i=0}^{M-1}$, treating the rotations $\{g_i\}_{i=0}^{M-1}$ as latent variables. 

To distinguish the model \eqref{eq:model_reconstruction} from \eqref{eq:model}, we highlight the key differences as follows:
\begin{itemize}
    \item[(i)] The parameter of interest is the unknown structure $V$ here, whereas it was the single rotation $g$ in model \eqref{eq:model};
    \item[(ii)] We observe $M$ i.i.d. samples instead of a single observation, meaning that all observed $M$ samples are used collectively to estimate the underlying volume structure $V$;
    \item[(iii)] Although the rotations $\{g_i\}_{i=0}^{M-1}$ are also unknown, they are treated as nuisance parameters, and we are not directly concerned with their estimation (though admittedly, more accurate estimation of $\{g_i\}_{i=0}^{M-1}$ could often contribute to better estimation of $V$).
\end{itemize}

The most common method for solving this reconstruction problem is the EM algorithm, which applies soft assignment iteratively, as outlined in Algorithm \ref{alg:volumeReconstruction}. In each iteration, the algorithm uses the volume estimate from the previous iteration, denoted as $\hat{V}^{(t)}$, to update the volume estimate $\hat{V}^{(t+1)}$, based on the observations $\ppp{y_i}_{i=0}^{M-1}$. The following proposition illustrates the relationship between the volume structure update rule at iteration $t+1$, and the MMSE orientation estimator introduced in \eqref{eqn:mmseEstimator}.

\begin{algorithm}[t!]
  \caption{\texttt{EM algorithm for volume reconstruction} \label{alg:volumeReconstruction}}

\textbf{Input:} An initial volume $\hat{V}^{(0)}$, number of iteration $T$ and  observations $\ppp{y_i}_{i=0}^{M-1}$ given by \eqref{eq:model_reconstruction}.\\
\textbf{Output:} Final volume estimation after $T$ iteration, $\hat{V}^{(T)}$.\\
\textbf{Each Iteration, for $t=0\ldots,T-1$:} %Given the last iterate $\hat{V}^{(t)}$, perform the following two steps:
\begin{enumerate}
    \item Compute for every $0 \leq i \leq M-1$:
        \begin{align}
            p_{i}^{(t)}(g) = \frac{\exp\Big(-\norm{y_i- \p{ g \circ \hat{V}^{(t)}}}^2/2\sigma^2\Big)}{ \int_{g\in\s{SO}(3)} \exp\Big(-\norm{y_i- \p{g \circ \hat{V}^{(t)}}}^2/2\sigma^2\Big)\der g},  \label{eqn:softMaxProbabilityDist}
        \end{align}
    \item Update the volume estimate:
        \begin{align}
            \hat{V}^{(t+1)} =&  \underset{V} {\argmin} \ \frac{1}{M} \sum_{i=0}^{M-1} \int_{g\in\s{SO}(3)} p_{i}^{(t)}(g) \norm{y_i - \p{ g \circ V}}^2 \der g\notag\\
            =& \frac{1}{M} \sum_{i=0}^{M-1} \int_{g\in\s{SO}(3)}  p_{i}^{(t)}(g) (g^{-1}\circ y_i) \der g.\label{eqn:emSoftAssignmentUpadteRule}
        \end{align}
\end{enumerate}
\end{algorithm}

To state the next result, we first introduce the \emph{MMSE back-rotation operator}. Given an observation $y_i$ and the current reference volume $\hat{V}^{(t)}$, let $p_i^{(t)}(g)$ denote the posterior weights over $g\in \s{SO}(3)$ as defined in~\eqref{eqn:softMaxProbabilityDist}. We define $\widehat{\mathfrak{g}}_{\s{MMSE},i,t}$ as the posterior average of the \emph{inverse action} $g^{-1}$:
\begin{align}
    \widehat{\mathfrak{g}}_{\s{MMSE},i,t} := \mathbb{E}\!\left[g^{-1}\,\big|\,y_i,\hat{V}^{(t)}\right],
    \label{eqn:g_MMSE_operator_def}
\end{align}
which should be interpreted as a linear operator acting on functions (or volumes). Namely, for any test function $f$ on which rotations act,
\begin{align}
    \big(\widehat{\mathfrak{g}}_{\s{MMSE},i,t}\circ f\big) := \int_{g\in \s{SO}(3)} p_i^{(t)}(g)\,\big(g^{-1}\circ f\big)\,\mathrm{d}g.
    \label{eqn:g_MMSE_operator_action}
\end{align}

\begin{proposition}[MMSE operator form of the EM M-step]
\label{thm:relationBetweenMMSEandEM}
Let $\hat{V}^{(t+1)}$ be the $(t+1)$-th volume estimator in the EM algorithm as described in Algorithm~\ref{alg:volumeReconstruction}, and let $\widehat{\mathfrak{g}}_{\s{MMSE},i,t}$ be defined as in~\eqref{eqn:g_MMSE_operator_def}--\eqref{eqn:g_MMSE_operator_action}. Then the M-step update can be written as
\begin{align}
    \hat{V}^{(t+1)} = \frac{1}{M}\sum_{i=0}^{M-1} \widehat{\mathfrak{g}}_{\s{MMSE},i,t}\circ y_i.
    \label{eqn:V_t_as_MMSE_est}
\end{align}
\end{proposition}

The proof of this proposition is presented in Appendix \ref{apx:relationBetweenMMSEandEM}. In words, the proposition shows that the update rule for the volume structure estimation at iteration $t+1$, given the volume  $\hat{V}^{(t)}$, is equivalent to aligning each observation $\ppp{y_i}_{i=0}^{M-1}$ using the associated back-rotation MMSE operator $\widehat{\mathfrak{g}}_{\text{MMSE}, i, t}$, computed based on the reference volume $\hat{V}^{(t)}$, and then averaging the aligned observations \eqref{eqn:V_t_as_MMSE_est}. Thus, the MMSE back-rotation operator is a key ingredient in the EM algorithm for volume reconstruction.

\begin{remark}
The \emph{MMSE back-rotation operator} $\widehat{\mathfrak{g}}_{\s{MMSE},i,t}$~\eqref{eqn:g_MMSE_operator_def} should be distinguished from the MMSE rotation \emph{matrix estimator} $\hat g_{\s{MMSE}}$ obtained in~\eqref{eqn:mmseEstimator} by taking a posterior mean in the $3\times 3$ matrix embedding and then projecting onto $\mathsf{SO}(3)$ (via Orthogonal Procrustes). 
In the matrix formulation, one forms $\hat g_{\s{relax}}$ as in~\eqref{eqn:relaxEstimator} and then projects to obtain a single $\mathsf{SO}(3)$ element. In contrast, in~\eqref{eqn:g_MMSE_operator_action} the inversion is handled \emph{inside} the posterior average: the expectation is taken over the inverse \emph{actions} $g^{-1}\circ(\cdot)$, yielding a linear operator that acts directly on $y_i$ (and is not, in general, itself a member of $\mathsf{SO}(3)$).
While these two objects differ in representation, they are both posterior-weighted aggregations over rotations; the operator form is the one that naturally appears in the EM M-step for reconstructing the volume.
\end{remark}

Despite the subtle differences arising from rotation estimation versus structure reconstruction, the previously introduced MMSE rotation estimator \eqref{eqn:mmseEstimator} can still serve as a practical tool for assessing the performance of soft assignments. Concretely, we take the posterior rotation matrix mean as in \eqref{eqn:g_MMSE_operator_def}, and then perform an Orthogonal Procrustes projection to replace $\widehat{\mathfrak{g}}_{\s{MMSE}, i, t}$. In this way, the MMSE rotation estimator provides a concrete approximation to the soft assignment in structure reconstruction. In numerical experiments, it is compared with the hard assignment, demonstrating superior performance relative to the maximum-likelihood estimator, as discussed below.

\paragraph{The MLE estimator as part of volume reconstruction.} 

In contrast to the soft assignment procedure, if we replace the MMSE operator with the corresponding MLE operator, we obtain the structure reconstruction algorithm by applying  \textit{hard assignment} iteratively. To be more specific, the MLE operator applied to the observations $\ppp{y_i}_{i=0}^{M-1}$ can be viewed as a hard assignment among all possible rotations. In practice, this procedure involves making a hard decision where a single rotation is selected from the rotation grid according to the closest alignment. In the $(t+1)$-th iteration, similarly to \eqref{eqn:V_t_as_MMSE_est}, the hard-assignment process can be expressed as follows:
\begin{align}
    \hat{V}^{(t+1)} = \frac{1}{M} \sum_{i=0}^{M-1} \hat{\mathfrak{g}}_{\s{MLE}, i,t}\circ y_i, \label{eqn:V_t_as_MAP_est}
\end{align}
where $\hat{\mathfrak{g}}_{\s{MLE}, i,t}$ is defined by
\begin{align}
    \hat{\mathfrak{g}}_{\s{MLE}, i, t} =& \argmax_{g^{-1}\in \s{SO}(3)} \P(g|y_i,\hat{V}^{(t)})\\
    =&\argmax_{g^{-1}\in \s{SO}(3)} p_i^{(t)}(g) =\argmin_{g^{-1}\in \s{SO}(3)}   \|y_i-( g \circ \hat{V}^{(t)})\|^2. 
\end{align}
In other words, $\hat{\mathfrak{g}}_{\s{MLE}, i, t} = (\hat{g}_{\s{MLE}, i, t})^{-1}$ where
\begin{align*}
\hat{g}_{\s{MLE}, i, t}:= \argmin_{g\in \s{SO}(3)}   \|y_i-( g \circ \hat{V}^{(t)})\|^2.  
\end{align*}
Hence, the hard assignment can be viewed as using the exact inverse of the MLE estimator introduced earlier.

\subsection{Empirical results for volume reconstruction and the ``Einstein from Noise'' phenomenon}\label{sec:empirical_reconstruction}
We demonstrate empirically volume reconstruction by applying the MLE estimator $\hat{g}_{\s{MLE}}$ and the MMSE estimator $\hat{g}_{\s{MMSE}}$ as part of the reconstruction problem, as specified in Section \ref{sec:connection}.

\begin{figure}[t!]
    \centering
    \includegraphics[width=1.0\linewidth]{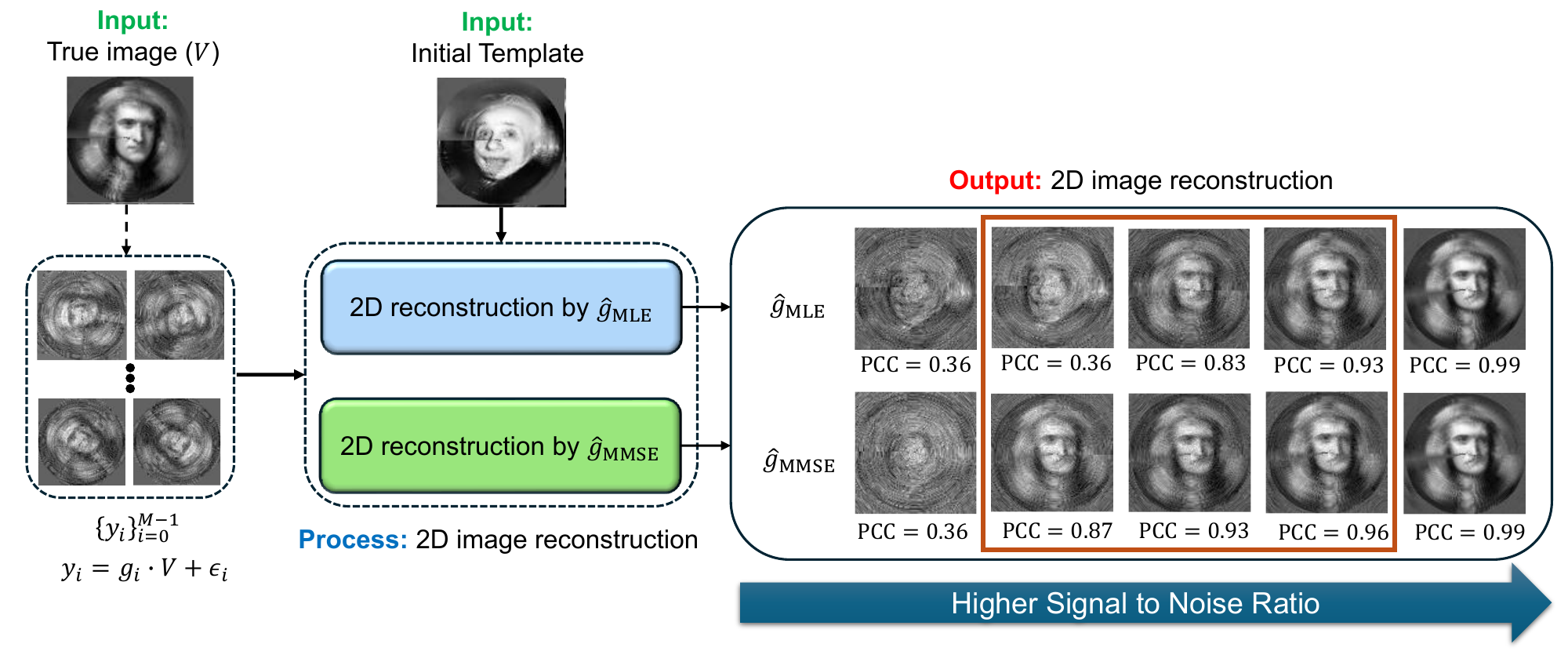}
    \caption{ \textbf{Comparison of 2D image recovery 
    using the MMSE and the MLE rotation estimators.} Iterative image recovery procedures with the MMSE estimator ($\hat{g}_{\text{MMSE}}$) and MLE estimator ($\hat{g}_{\text{MLE}}$) are defined in \eqref{eqn:V_t_as_MMSE_est} and \eqref{eqn:V_t_as_MAP_est}, respectively. The experiment employs a template image of Einstein and a ground truth image of Newton, both rotated in 2D over a uniform polar grid with $L = 30$ samples. Each image is of size $100\times 100$ pixels, and the radial direction is discretized using $R=300$ points. The reconstructed images within the dark-orange rectangle (right panel) show superior performance with the MMSE rotation estimator, with Pearson cross-correlation (PCC) provided for each reconstructed image. The MLE and MMSE reconstructions are nearly identical at high SNR ($\sigma \to 0$), as predicted by Proposition \ref{thm:mapAndMmseEstimatorsConicideness}. The SNR values used for the panels (from right to left) are $10^{-2}$, $4 \times 10^{-3}$, $2 \times 10^{-3}$, $7 \times 10^{-4}$, and $2 \times 10^{-4}$. At very low SNR ($\sigma \to \infty$), the ``Einstein from Noise'' effect appears, where the estimator resembles the template image of Einstein rather than the underlying truth of Newton. In intermediate SNR ranges, using the MMSE estimator in the iterative step clearly outperforms the MLE estimator. 
    }
    \label{fig:4}
\end{figure}

\begin{figure}[t!]
    \centering
    \includegraphics[width=1.0\linewidth]{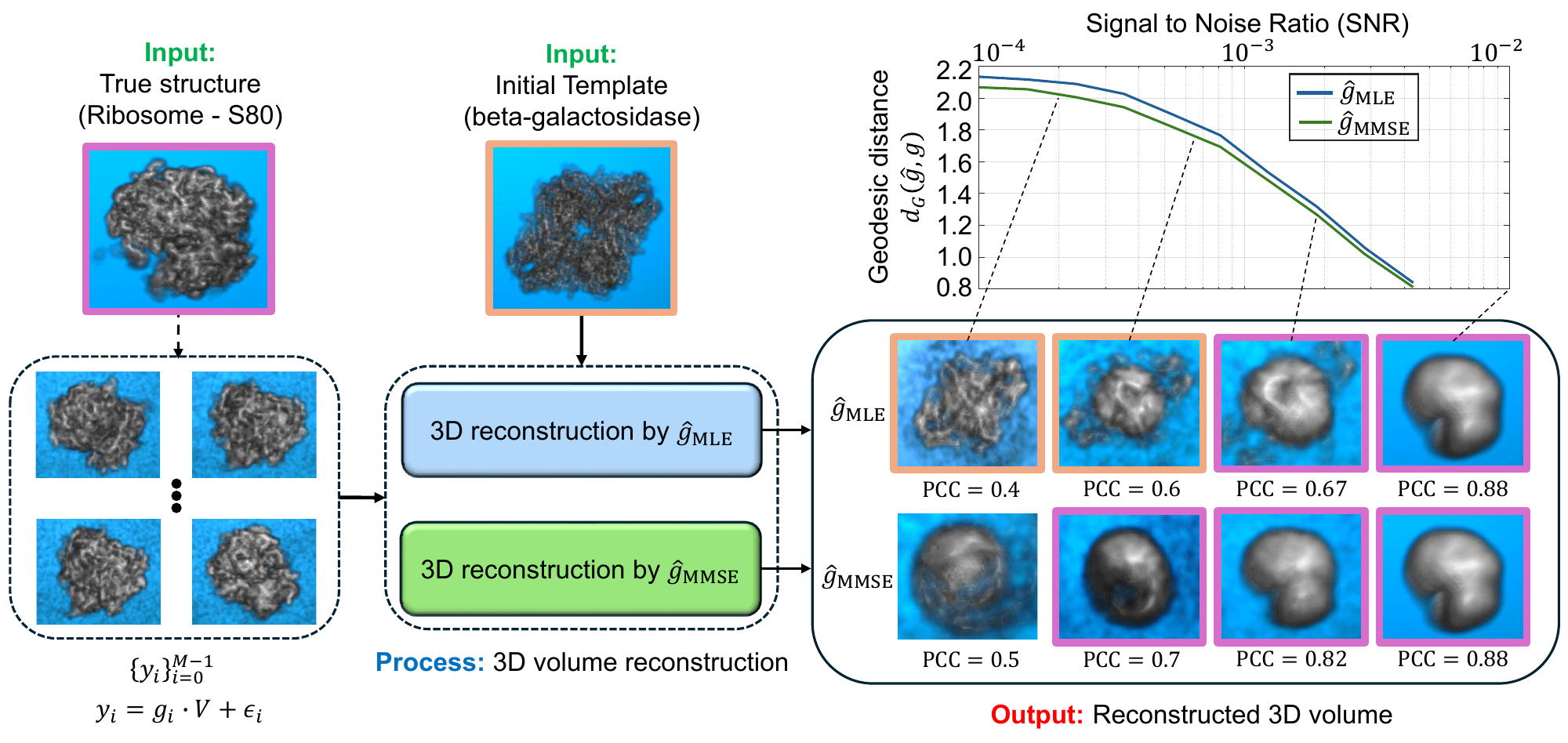}
    \caption{{\textbf{Comparison of 3D structure reconstruction in cryo-ET subtomograms averaging using the MMSE and MLE rotation estimators.} Iterative structure reconstruction procedures with the MMSE estimator ($\hat{g}_{\text{MMSE}}$) and MLE estimator ($\hat{g}_{\text{MLE}}$) are defined in \eqref{eqn:V_t_as_MMSE_est} and \eqref{eqn:V_t_as_MAP_est}, respectively. At high SNR levels, a low-resolution structure emerges due to finite grid sampling of the rotation group $\s{SO}(3)$, effectively acting as a low-pass filter. The 3D reconstruction using the MMSE estimator consistently outperforms the reconstruction using the MLE estimator. For high SNR conditions (i.e., $\sigma \to 0$), both estimators yield similar 3D structures, as expected from Proposition \ref{thm:mapAndMmseEstimatorsConicideness}. The SNR values used for the panels (from right to left) are $10^{-2}$, $2 \times 10^{-3}$, $7 \times 10^{-4}$, and $2 \times 10^{-4}$, with a volume size of $32 \times 32 \times 32$. The boxes highlighted in purple resemble the true input structure (Ribosome-S80~\cite{wong2014cryo}), while the orange-highlighted boxes are more similar to the initial template (beta-galactosidase~\cite{bartesaghi2014structure}), illustrating the ``Einstein from Noise'' phenomenon. Notably, at very low SNRs, the ``Einstein from Noise" effect is evident with the MLE estimator but not with the MMSE estimator.}}    
    \label{fig:5}
\end{figure}

\paragraph{Description of the experiments.}
We demonstrate the reconstruction processes, which integrates MLE and MMSE rotation estimators as intermediate steps, using the two examples: 2D image recovery (Figure \ref{fig:4}) and 3D volume reconstruction without projection (Figure \ref{fig:5}). The iterative reconstruction process, as outlined in Algorithm \ref{alg:volumeReconstruction}, was performed until convergence (i.e., when the relative difference between consecutive iterations fell below a predefined threshold of $10^{-3}$ or until reaching a maximum of 100 iterations.

Figures \ref{fig:4} and \ref{fig:5} were generated using slightly different methods. 
For the 2D experiment presented in Figure \ref{fig:4}, we used polar coordinates, while in the 3D experiment, shown in Figure \ref{fig:5}, we used a standard Cartesian basis. The primary difference lies in the interpolation required for producing Figure \ref{fig:5}, which utilizes cubic interpolation for each observation based on the estimated rotation. This introduces certain ``quantization" errors. 
Additionally, as detailed in Section \ref{sec:problemFormulation}, the 3D reconstruction process, presented in Figure \ref{fig:5}, requires a ``rounding'' step which amounts to solving the Orthogonal Procrustes procedure. 
In Figure \ref{fig:4}, the true and template structures were generated in a polar representation with $d = 300$ radial points and $L = 30$ polar angle points. The reconstruction process was performed with $M = 5 \times 10^4$ observations. The additive noise was added in the polar representation. In Figure \ref{fig:5}, $M = 3000$ observations were used, with the rotation group $\s{SO}(3)$ grid size of $L = 300$, and a volume size of $32 \times 32 \times 32$. 

\paragraph{Empirical observations.}
A few observations can be made from  Figures \ref{fig:4} and \ref{fig:5}. First, in the case of high SNR (i.e., as $\sigma \to 0$), the volume reconstruction is similar whether using the MLE estimator or the MMSE estimator. This similarity is theoretically supported by Proposition~\ref{thm:mapAndMmseEstimatorsConicideness}. However, as the SNR decreases, the reconstructions diverge, with the volumes reconstructed using the MMSE estimator showing a better correlation with the true volume.
Second, in scenarios of extremely low SNR, where the structural signal is nearly nonexistent, the phenomenon known as ``Einstein from Noise'' manifests in both 2D and 3D contexts. This phenomenon pertains to the inherent model bias within the reconstruction procedure, specifically in relation to the initial templates. In such cases, the reconstructed volume exhibits structural similarities to the initial template, even though the observations do not substantiate this outcome. The generation of a structured image from entirely noisy data has attracted considerable attention, particularly during a significant scientific debate regarding the structure of an HIV molecule~\cite{mao2013molecular,henderson2013avoiding,van2013finding,subramaniam2013structure,mao2013reply}; for a comprehensive description and statistical analysis, see~\cite{balanov2024einstein, balanov2024confirmation,balanov2025expectation,balanov2025structure}. Notably, our empirical evidence suggests that the ``Einstein from Noise'' phenomenon is more pronounced when adopting the MLE estimator compared to the MMSE estimator, implying that the MMSE approach is less vulnerable to the choice of the initial template. Furthermore, our experiment suggests that the advantage of using the Bayesian MMSE estimator over the MLE estimator is more significant in 3D structure reconstruction tasks compared to 2D image recovery, where the 3D setting is a problem of greater interest to researchers in structural biology.

\section{Structural heterogeneity analysis}
\label{sec:heterogeneityAnalysis}

Understanding heterogeneity is central to revealing the dynamic behavior of macromolecular complexes in cryo-EM. Unlike traditional 3D classification methods that assume a small number of discrete states, modern approaches aim to capture continuous structural variability by embedding projection images into a latent conformational space.
Here, we adopt the \textsc{RECOVAR} framework~\cite{gilles2025cryo}, which performs heterogeneity analysis using principal component analysis (PCA) based on a regularized estimate of the conformational covariance matrix. The key idea is to infer the covariance of 3D volumes directly from noisy 2D projection images, leveraging the fact that although the volumes themselves are unobserved, their projections contain sufficient statistical information.

\subsection{Fixed-pose methods}

A central computational assumption in \textsc{RECOVAR} is that the poses $\{\phi_i\}$ of all particle images are known and fixed, typically obtained from a consensus refinement procedure. This ``fixed-pose'' assumption underlies many modern heterogeneity analysis pipelines, including 3D Classification in \textsc{cryoSPARC}~\cite{punjani2017cryosparc}, 3D Variability Analysis (3DVA)~\cite{punjani20213d}, 3DFlex~\cite{punjani20233dflex}, \textsc{cryoDRGN}~\cite{zhong2021cryodrgn}, \textsc{CryoDRGN-AI-fixed}~\cite{levy2024end}, \textsc{Opus-DSD}~\cite{luo2023opus}, and \textsc{RECOVAR}~\cite{gilles2025cryo}. These methods are collectively referred to as \emph{fixed-pose methods}, since they treat poses as known inputs during downstream inference. 
Among them, \textsc{RECOVAR} has recently emerged as one of the most effective approaches. According to the \textsc{CryoBench} evaluation~\cite{jeon2024cryobench}, which benchmarks multiple methods across diverse synthetic and experimental datasets, \textsc{RECOVAR} consistently achieves state-of-the-art performance in both structural resolution and latent space recovery. Notably, \textsc{CryoBench} evaluations were performed by external users, indicating \textsc{RECOVAR}’s strong out-of-the-box performance and robustness to hyperparameter choices. This motivates our choice to build upon the \textsc{RECOVAR} framework. 

Despite its strong empirical performance, the fixed-pose assumption merits careful scrutiny. In practice, ground-truth poses are never directly accessible and cannot be recovered exactly, even if the underlying volume is known, due to noise and the intrinsic ill-posedness of the inverse problem. Most existing methods address this by using MLE pose estimates in real data. This gap between assumed and achievable pose accuracy motivates a central question: to what extent can more accurate pose estimation improve downstream heterogeneity recovery? In this work, we address this question by replacing MLE poses with our MMSE estimator within \textsc{RECOVAR}.

\subsection{The mathematical model}

The forward model \cite[Eq.~(1)]{gilles2025cryo} used in \textsc{RECOVAR} is expressed in the Fourier domain as
\begin{align}
    \tilde{y}_i = C_i \hat{P}(\phi_i) \tilde{V}_i + \tilde{\varepsilon}_i, \label{eqn:observationFourierDomain}
\end{align}
where $\tilde{y}_i \in \mathbb{C}^{N^2}$ is the observed Fourier-transformed image  (with $N$ denoting the grid size along each spatial axis), $\tilde{V}_i \in \mathbb{C}^{N^3}$ is the (unknown) 3D Fourier volume corresponding to a particular conformation, $\hat{P}(\phi_i)$ is the tomographic projection operator from 3D to 2D after a rigid-body motion $\phi_i = (g_i, t_i)$ with $g_i \in \mathsf{SO}(3)$ a rotation and $t_i \in \mathbb{R}^2$ an in-plane shift, $C_i$ is the contrast transfer function (CTF), and $\tilde{\varepsilon}_i$ is additive noise.

This formulation is a discretized and sampled version of our continuous forward model in~\eqref{eq:cryoEMfull_Fourier}. Specifically, the operator $\hat{P}(\phi_i)$ acting on $\tilde{V}_i$ corresponds directly to $e^{-2\pi i \boldsymbol{\omega} \cdot t_i} \cdot (g_i \circ \tilde{V})(\omega_1, \omega_2, 0)$, where each component reflects the same physical process, namely, in-plane shift, 3D rotation, and evaluation on the central slice in Fourier space.

To model structural variability, we treat each underlying volume $\tilde{V}_i$ as a random sample from a distribution over 3D conformations. We assume this distribution has a well-defined mean $\mu \in \mathbb{C}^{N^3}$ and covariance $\Sigma \in \mathbb{C}^{N^3 \times N^3}$ given by
\begin{align}
    \mu = \mathbb{E}[\tilde{V}], \quad \Sigma = \mathbb{E}[(\tilde{V} - \mu)(\tilde{V} - \mu)^*],
\end{align}
where the expectation is taken over the conformational distribution. The objective of \textsc{RECOVAR} is to estimate both $\mu$ and $\Sigma$ from the observed projections ${\tilde{y}_i}$ with given poses ${\phi_i}$, using regularized least-squares minimization~\cite[Eqns.~(2)–(3)]{gilles2025cryo}.

\subsection{Numerical experiments}

\begin{figure}[t!]
    \centering
    \includegraphics[width=0.9\linewidth]{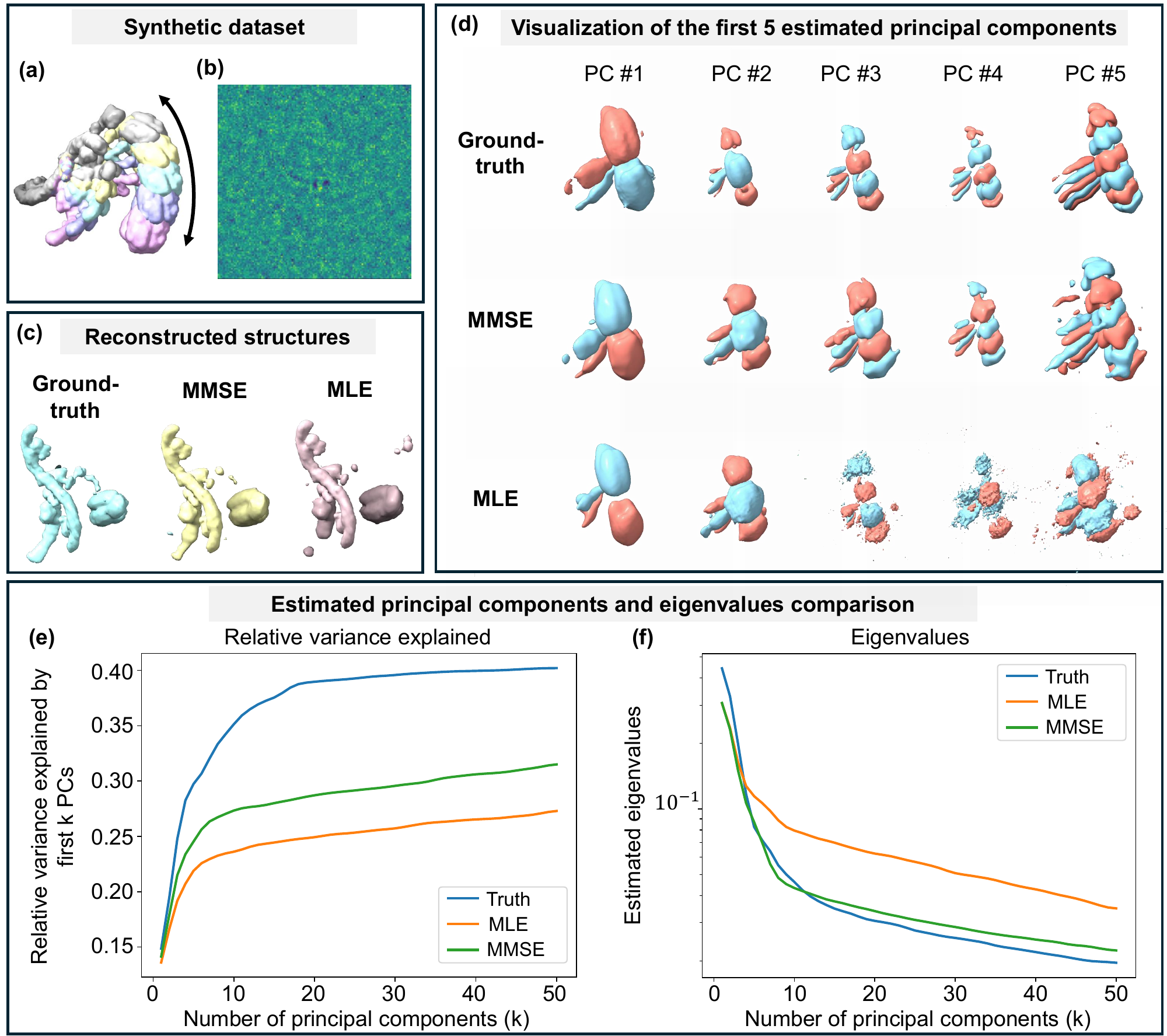}
    \caption{\textbf{MMSE pose estimation improves heterogeneity reconstruction over MLE.} 
    (\textbf{a}) Ground-truth density maps from a synthetic dataset simulating a one-dimensional conformational transition, a standard benchmark for methods such as \textsc{cryoDRGN} \cite{zhong2021cryodrgn} and \textsc{RECOVAR} \cite{gilles2025cryo}. Fifty equally spaced states approximate the continuous trajectory, with colors indicating positions along the pathway (only five states are shown in the figure for clarity). 
    (\textbf{b}) Example of a noisy projection image from the high-noise regime used in this study, where even ground-truth poses recover only $\sim 40\%$ of the variance for the top 30 components due to finite-sample and high-noise effects. 
    (\textbf{c}) Comparison of reconstructed structures using different pose priors (ground-truth, MMSE, MLE), showing MMSE yields structures closer to the true conformation state.
    (\textbf{d}) First five principal components estimated using ground-truth poses, MMSE pose estimates, and MLE pose estimates. MMSE results closely match the ground truth in structural detail, whereas MLE reconstructions degrade beyond the first modes, becoming blurrier and less representative of the underlying variability. 
    (\textbf{e}) Subspace accuracy, measured as the percentage of total variance captured (as defined in \eqref{eqn:totalVarianceCaptured}). Ground truth achieves $\sim 40\%$, MMSE $\sim 30\%$, and MLE $\sim 25\%$. 
    (\textbf{f}) Eigenvalue recovery: both methods estimate the largest true eigenvalues well, but MMSE remains accurate for smaller eigenvalues, while MLE substantially overestimates them. These results show that MMSE pose estimation consistently yields more accurate recovery of conformational variability than MLE.}
    \label{fig:6}
\end{figure}

In our experiments, we consider a synthetic one-dimensional \textit{conformational transition}, a standard benchmark for continuous heterogeneity analysis methods such as \textsc{cryoDRGN} \cite{zhong2021cryodrgn} and \textsc{RECOVAR} \cite{gilles2025cryo}. The corresponding ground-truth density maps along the conformational coordinate are shown in Figure~\ref{fig:6}(a). To generate the projection images, the underlying conformational states were uniformly sampled along the one-dimensional transition, and molecular orientations were sampled from a uniform distribution over $\mathsf{SO}(3)$. Figure~\ref{fig:6}(b) further displays a representative noisy projection under high-noise conditions. 

To avoid storing and manipulating the full high-dimensional covariance matrix $\Sigma \in \mathbb{C}^{N^3 \times N^3}$, \textsc{RECOVAR} first estimates a low-dimensional subspace of rank $r \ll N^3$ in which the covariance is approximated. Following~\cite[Appendix~A.2]{gilles2025cryo}, we estimate $\hat{\Sigma}$ via regularized least-squares, then form a rank-$d$ approximation $\hat{\Sigma}_{\mathrm{col}}$ by selecting a subset of columns using a greedy SNR-based criterion that ensures both high SNR and low inter-column correlation. An orthonormal basis $\tilde{U} \in \mathbb{C}^{N^3 \times r}$ spanning $\hat{\Sigma}_{\mathrm{col}}$ is obtained via randomized SVD. The conformational mean $\hat{\mu}$ is also estimated at this stage.
Each conformational volume is then represented as
\begin{align}\label{eq:embedding}
    \tilde{V}_i = \tilde{U} z_i + \hat{\mu},
\end{align}
where $\hat{\mu}$ is the estimated conformational mean from the first stage, and $z_i \in \mathbb{C}^r$ are low-dimensional coordinates. In \textsc{RECOVAR}, these coordinates are not explicitly recovered; their covariance matrix $\hat{\Sigma}_{\tilde{U}} \in \mathbb{C}^{r \times r}$ is estimated directly via a reduced least-squares problem (Eq.~(14) in~\cite{gilles2025cryo}), yielding a low-rank covariance approximation that characterizes the variability of $z_i$. The eigen-decomposition of $\hat{\Sigma}_{\tilde{U}}$ then provides the estimated principal components and eigenvalues in the reduced space. The complete procedure is detailed in~\cite[Algorithm~1]{gilles2025cryo}.

We perform our analyses on a controlled synthetic dataset of $30{,}000$ projection images ($N = 128$), simulated from a one-dimensional conformational transition discretized into $50$ equally spaced states to approximate continuous heterogeneity~\cite{gilles2025cryo}. We first evaluate the effect of orientation estimation on covariance recovery. The study is conducted in the high-noise regime of \textsc{RECOVAR} (Figure~\ref{fig:6}(b)), where even with ground-truth poses, the top $30$ principal components capture only about $40\%$ of the variance due to finite-sample and high-noise effects. We set $r= 50$ and compare three pose sources: ground truth, MMSE estimates, and MLE estimates, keeping all other steps identical. For fixed-pose methods, poses are treated as given, so computational cost remains unchanged.

The accuracy of the estimated subspace is quantified by the percentage of total variance captured:
\[
\|U_k^* \Sigma U_k\|_S / \|\Sigma\|_S, \label{eqn:totalVarianceCaptured}
\]
where $U_k$ contains the first $k$ estimated principal components, $\Sigma$ is the ground-truth covariance matrix, and $\Sigma^{1/2}$ is its matrix square root. Here, $\|A\|_S = \sum \sigma_i(A)$ is the Schatten $1$-norm of matrix $A$ with $\sigma_i$ denoting the singular values of $A$.

Figures~\ref{fig:6}(e) and~\ref{fig:6}(f) present the variance-capture results as described and the corresponding eigenvalue recovery, respectively. Figure~\ref{fig:6}(d) visualizes the first five recovered principal components. Overall, MMSE pose estimation yields principal components and eigenvalue spectra that are consistently closer to the ground truth than those obtained from MLE pose estimation.

We further perform a more granular evaluation of structure reconstruction at the level of individual conformational states. For this analysis, we select all projection images corresponding to a specific conformational state and embed them into the latent conformational space ($z$-space, see \eqref{eq:embedding}) using \textsc{RECOVAR}, conditioned on different pose priors: ground-truth, MMSE estimates, and MLE estimates. For each subset, the reconstruction corresponding to the mean embedding serves as an estimate of that specific conformational state. These reconstructions provide a direct measure of how different orientation estimates affect the accuracy of heterogeneity structure reconstruction. Figure~\ref{fig:6}(c) demonstrates that MMSE orientation estimates produce reconstructed structures closer to the ground-truth conformation than those obtained from MLE, further highlighting the advantage of MMSE pose estimation in high-noise heterogeneity analysis. We further compared the reconstructions obtained using MMSE and MLE orientation estimates against the ground-truth conformational state. The reported local FSC resolution scores (in~\text{\AA}) show that the 50\% quantile resolutions are 11.45~\text{\AA} and 11.55~\text{\AA} for MMSE and MLE, respectively, while the 90\% quantiles are 18.83~\text{\AA} and 21.43~\text{\AA} (also see \cite{gilles2025cryo} Supplementary S.J.).

\section{Discussion and conclusions} \label{sec:conclutions}

In this work, we have introduced the Bayesian framework for enhancing orientation estimation for various applications in structural biology. The proposed approach offers greater flexibility and improved accuracy compared to existing methods, with the MMSE estimator as a prime example. This technique handles diverse structural conformations and arbitrary rotation distributions across sample sets. Our empirical results establish that the proposed MMSE estimator consistently surpasses the performance of the current methods, particularly in challenging low SNR environments, as well as when prior information on rotation distribution is available or approximately known. We provide a theoretical foundation to explain these performance gains. As rotation determination is crucial for both 2D and 3D reconstruction processes, we further illustrate how utilizing the MMSE estimator as a soft-assignment step in iterative refinement leads to significant improvements over hard-assignment methods. Moreover, the proposed Bayesian approach empirically offers enhanced resilience against the ``Einstein from Noise" phenomenon, effectively reducing model bias and improving the overall reliability of structural reconstructions. 
Thus, our main recommendation is to adopt the Bayesian MMSE rotation estimator over the MLE estimator in related application scenarios. Already integrated into most software, the Bayesian rotation estimator can be easily implemented with minimal computational cost, offering improved accuracy and resilience for orientation determination and structure reconstruction.

\paragraph{Future work.} The MMSE orientation estimator represents the most natural Bayes estimator in the context of cryo-EM, as it coincides with the M-step reconstruction of the EM algorithm. Nevertheless, alternative Bayes estimators associated with different loss functions could also be explored in future studies.
A particularly promising direction is the direct estimation of the rotation distribution from observed images and the incorporation of this information into rotation estimation or EM-based 3D reconstruction algorithms~\cite{janco2022accelerated, xu2025misspecified}. Such approaches could substantially improve rotation accuracy, especially in low-SNR regimes, as indicated by our results in Figure~\ref{fig:2}, and consequently enhance volume reconstruction quality.
Further extensions may involve modeling structural uncertainties, particularly relevant for flexible or heterogeneous proteins, and addressing general pose estimation problems that jointly consider rotations and translations. Finally, leveraging the Bayesian framework to derive confidence regions for individual rotations presents an exciting opportunity to improve the interpretability and reliability of rotation estimation in cryo-EM.

\section*{Data Availability}
The detailed implementation and code are available at \href{https://github.com/AmnonBa/bayesian-orientation-estimation}{https://github.com/AmnonBa/bayesian-orientation-estimation}.

\section*{Acknowledgment}
A.S. and S.X. are supported in part by AFOSR under Grant FA9550-23-1-0249, the Simons Foundation Math+X Investigator Award, NSF under Grant DMS 2510039, and NIH/NIGMS under Grant R01GM136780-01. T.B. is supported in part by BSF under Grant 2020159, in part by NSF-BSF under Grant 2024791, in part by ISF under Grant 1924/21, and in part by a grant from The Center for AI and Data Science at Tel Aviv University (TAD). We thank Marc Aur\'{e}le Gilles for providing guidance on \textsc{RECOVAR} software. We also thank Marc Aur\'{e}le Gilles, Eric Verbeke, and Ruiyi Yang for their helpful discussions. 

\bibliographystyle{plain}
\bibliography{main_arxiv}

\begin{appendices}

{\centering{\section*{Appendix}}}

\section{Full cryo-EM model} \label{appx:fullCryoEmModel}

The (homogeneous) cryo-EM imaging process can be described by a comprehensive generative model that incorporates the key physical components of image formation, including random orientations, in-plane shifts, tomographic projection, and contrast transfer function (CTF).

Formally, each observed 2D image $y_i$ is modeled as
\begin{equation} \label{eq:cryoEMfull}
y_i(x_1,x_2) = h_i * T_{t_i} \Big(\Pi \big((g_i \circ V)(x_1,x_2,x_3)\big)\Big) + \varepsilon_i(x_1,x_2),~~~\text{for }(x_1,x_2,x_3)\in\R^3,
\end{equation}
where $V$ denotes the underlying 3D volume, $g_i$ denotes the random rotation, $\Pi$ is the tomographic projection operator (i.e., line integration along $x_3$ axis), $T_{t_i}$ denotes the in-plane shift operator with shift $t_i\in\R^2$, $*$ denotes the 2D convolution in the image plane, $h_i$ is the point spread function (whose Fourier transform is commonly known as CTF), and $\varepsilon_i$ denotes the measurement noise. Equivalently, by applying the Fourier transform and using the Fourier Slice Theorem, we obtain the forward model in the Fourier domain:
\begin{equation} \label{eq:cryoEMfull_Fourier}
\tilde{y}_i(\omega_1,\omega_2) = \tilde{h}_i\cdot e^{-2\pi\i \boldsymbol{\omega}\cdot t_i} \cdot \p{(g_i \circ \tilde{V})(\omega_1,\omega_2,0)} + \tilde\varepsilon_i(\omega_1,\omega_2),~~~\text{for }\boldsymbol{\omega}=(\omega_1,\omega_2)\in\R^2,
\end{equation}
where $\tilde{y}_i,\tilde{h}_i,\tilde\varepsilon_i$ are the 2D Fourier transforms of $y_i,h_i,\varepsilon_i$, respectively, and $\tilde{V}$ is the 3D Fourier transform of $V$.

In practice, both the images and the underlying volume are discretized and sampled on finite grids. Depending on whether the model is implemented in the spatial or Fourier domain, different but equivalent discretized versions arise. Specifically:
\begin{itemize}
    \item In the \textbf{spatial domain}, each image \( y_i \) is represented as a vector in \( \R^{N^2} \), and the 3D volume \( V \) is discretized as an array in \( \R^{N^3} \).
    \item In the \textbf{Fourier domain}, we work with complex-valued data, where \( \tilde{y}_i \in \mathbb{C}^{N^2} \) and \( \tilde{V} \in \mathbb{C}^{N^3} \) are the discrete Fourier representations of the image and volume, respectively. Since the images and the volume are real-valued, it follows that $\tilde{y}_i$ and $\tilde{V}$ are conjugate symmetric, i.e., $\overline{\tilde{y}_i(\omega_1,\omega_2)} = \tilde{y}_i(-\omega_1,-\omega_2)$ and $\overline{\tilde{V}(\omega_1,\omega_2,\omega_3)} = \tilde{V}(-\omega_1,-\omega_2,-\omega_3)$.
\end{itemize}
Here, \( N \) denotes the resolution of the reconstructed grid: each image consists of \( N \times N \) pixels, and the volume is represented on an \( N \times N \times N \) voxel grid.

In the heterogeneous case, the underlying volume $V$ may itself be a realization drawn from a distribution over conformational states $V_i\sim \mathcal{P}_V$ where $\mathcal{P}_V$ represents a (possibly low-dimensional) distribution capturing the variability in molecular conformations, such as discrete states, continuous manifolds. This setting gives rise to the heterogeneous cryo-EM problem, where the goal is to infer not just a single structure, but the distribution $\mathcal{P}_V$ from noisy 2D projections.

\section{Bayes pose estimation with rotations and shifts}
\label{apx:mmse_rot_shift}

To account for in-plane shifts, we model the latent pose as $\phi=(g,t)\in \mathsf{SO}(3)\times\mathbb{R}^2$, where $g$ denotes the 3D rotation and $t$ denotes the in-plane translation. Given a volume $V$, the forward model can be written as
\begin{align}
    p\!\left(y \mid g,t\right) \propto \exp\!\Big(-\tfrac{1}{2}\,\|y-\mathcal{A}(g,t)V\|_{\Sigma^{-1}}^2\Big),
    \label{eq:mmse_pose_likelihood}
\end{align}
where $\mathcal{A}(g,t)$ is the cryo-EM/ET forward operator including both rotation and shift (e.g., in the Fourier domain the shift acts via the phase factor $e^{-2\pi i\,\omega\cdot t}$; cf.\ Appendix~\ref{appx:fullCryoEmModel}).
We assume priors $\Lambda(g)$ on rotations and $\pi(t)$ on shifts. The joint posterior then takes the form
\begin{align}
    p(g,t\mid y) \propto p(y\mid g,t)\,\Lambda(g)\,\pi(t).
    \label{eq:mmse_pose_posterior}
\end{align}

\paragraph{Additive loss function.}
To define a Bayes estimator, we specify a loss function on $\mathsf{SO}(3)\times\mathbb{R}^2$. Throughout, we adopt an \emph{additive} pose loss,
\begin{align}
    \mathcal{L}\big((g,t),(\widehat{g},\widehat{t}\,)\big) = \mathcal{L}_{g}(g,\widehat{g} \,) + \mathcal{L}_t(t,\widehat{t} \,),
\label{eq:mmse_pose_additive_loss}
\end{align}
which penalizes rotation and translation errors separately (optionally with application-dependent weights to reconcile units).
Under~\eqref{eq:mmse_pose_additive_loss}, the Bayes pose estimator is defined by
\begin{align}
    (\widehat{g}_{\mathrm{Bayes}},\widehat t_{\mathrm{Bayes}}) \in
    \argmin_{\widehat{g}\in\mathsf{SO}(3),\,\widehat{t}\in\mathbb{R}^2} \ \mathbb{E}\!\left[\mathcal{L}\big((g,t),(\widehat{g},\widehat t \,)\big)\,\big|\,y\right],
\label{eq:mmse_pose_definition}
\end{align}
where the expectation is taken with respect to the \emph{joint} posterior~\eqref{eq:mmse_pose_posterior}.
Because the loss is additive, the minimization in~\eqref{eq:mmse_pose_definition} decomposes into two marginal Bayes estimators:
\begin{align}
    \widehat{g}_{\mathrm{Bayes}}
    & =  \argmin_{\widehat{g}\in\mathsf{SO}(3)} \ \mathbb{E}\!\left[\mathcal{L}_{g}(g,\widehat{g})\mid y\right],
    \label{eq:mmse_rot_marginal}\\
    \widehat{t}_{\mathrm{Bayes}} & = \argmin_{\widehat t\in\mathbb{R}^2} \ \mathbb{E}\!\left[\mathcal{L}_{t}(t,\widehat{t}\,)\mid y\right].
\label{eq:mmse_shift_marginal}
\end{align}

\paragraph{Closed-form MMSE rules under squared loss.}
For translations, choosing Euclidean squared loss $\mathcal{L}_t(t,\widehat{t}\,)=\|t-\widehat{t}\,\|_2^2$ yields the posterior mean (similarly to~\eqref{eqn:relaxEstimator})
\begin{align}
    \widehat{t}_{\mathrm{MMSE}} = \mathbb{E}[t\mid y] = \int_{\mathsf{SO}(3)}\int_{\mathbb{R}^2} t\,p(g,t\mid y)\,\der t\,\der g,
    \label{eq:mmse_shift_posterior_mean}
\end{align}
where $p(g,t\mid y)$ is defined through~\eqref{eq:mmse_pose_posterior}.
For rotations, when $g$ is represented by its $3\times 3$ matrix and we take the squared chordal/Frobenius loss
$\mathcal{L}_{g}(g,\widehat{g})=\|g-\widehat{g}\|_F^2$,
the corresponding Bayes estimator is obtained by first forming the relaxed posterior mean matrix (similarly to~\eqref{eqn:relaxEstimator})
\begin{align}
    \widehat{g}_{\mathrm{relax}}  = \mathbb{E}[g \mid y] = \int_{\mathsf{SO}(3)}\int_{\mathbb{R}^2} g\,p(g,t\mid y)\,\der t\,\der g,
    \label{eq:mmse_rot_relax}
\end{align}
and then projecting it onto $\mathsf{SO}(3)$ via the Orthogonal Procrustes map (similarly to~\eqref{eqn:propOrthProc}),
\begin{align}
    \widehat{g}_{\mathrm{MMSE}} = \argmin_{\Omega\in \mathsf{SO}(3)} \ \big\|\widehat{g}_{\mathrm{relax}}-\Omega\big\|_F .
    \label{eq:mmse_rot_procrustes}
\end{align}
which returns the closest valid rotation matrix (in Frobenius norm) to $\widehat{g}_{\mathrm{relax}} $.

\section{Alignment between two noisy volumes} \label{appx:alignmnetBetweenNoisyVolumes}
The model described in \eqref{eq:model} can be generalized to address the alignment problem between two noisy volumes. In \eqref{eq:model}, a single observation $y$ was assumed to follow the model, with the volume $V$ considered deterministic and known. For aligning two noisy volumes, we now consider two observations $y_i$ for $i = 1, 2$, given by:
\begin{align}
    y_i = g_i \circ V + \epsilon_i, \label{eqn:app_A1}
\end{align}
where $\epsilon_i \sim \mathcal{N}(0, \Sigma_{d \times d})$ for a positive-definite cocaraince matrix $\Sigma_{d \times d}$, and $g_i$ are drawn from a distribution $\Lambda$ over $\text{SO}(3)$. We assume that $\epsilon_1, \epsilon_2, g_1, g_2$ are independent, and that the volume $V$ is deterministic but unknown. Furthermore, we assume that the noise statistics are invariant under the group actions of $\text{SO}(3)$, meaning the distributions of $\epsilon$ and $g \circ \epsilon$ are identical for any $g \in \text{SO}(3)$. The objective is to determine the ``best'' alignment $\hat{g} \in \text{SO}(3)$ between $y_1$ and $y_2$, according to criteria such as the MLE, MAP, or MMSE estimators.

From \eqref{eqn:app_A1}, the unknown volume $V$ can be expressed as:
\begin{align}
    V = g_1^{-1} \circ (y_1 - \epsilon_1). \label{eqn:app_A2}
\end{align}
Substituting \eqref{eqn:app_A2} into \eqref{eqn:app_A1}, yields:
\begin{align}
    y_2 &= g_2 \circ V + \epsilon_2 \\
    &= g_2 \circ \left(g_1^{-1} \circ (y_1 - \epsilon_1)\right) + \epsilon_2 \\
    &= (g_2 \cdot g_1^{-1}) \circ y_1 - (g_2 \cdot g_1^{-1}) \circ \epsilon_1 + \epsilon_2 \label{eqn:app_A5}
    \\&= (g_2 \cdot g_1^{-1}) \circ y_1 + \epsilon, \label{eqn:app_A6}
\end{align}
where \eqref{eqn:app_A5} follows from the associativity of group actions, and in\eqref{eqn:app_A6}, we have used $\epsilon \triangleq (g_2 \cdot g_1^{-1}) \circ \epsilon_1 + \epsilon_2$. Since the noise statistics are invariant under group actions, it follows that $\epsilon \sim \mathcal{N}(0, 2\Sigma_{d \times d})$.

Thus, the problem of estimating the relative rotation $g_2 \cdot g_1^{-1}$ between $y_1$ and $y_2$ is similar to estimating $y_2$ under the assumption that $V$ is known. However, the covariance of the additive noise $\epsilon$ in this estimation is doubled compared to the case when $V$ is known.

\section{Proof of Proposition \ref{thm:orthogonalProcrustes}} \label{appx:proofOrthogonalProcrustes}
By the definition of $\hat{g}_{\s{MMSE}}$ \eqref{eqn:mmseEstimator}, we have,
\begin{align}
    \hat{g}_{\s{MMSE}} = \underset{\hat{g} \in \s{SO}(3)} {\argmin} \int_{\s{SO(3)}} \norm{g - \hat{g}}_F^2 \P_\Lambda(g|y, \barV)\der g.
\end{align}
Then, as $\norm{g}_F, \norm{\hat{g}}_F$ are fixed for any $g, \hat{g} \in \s{SO}(3)$, we have,
\begin{align}
    \hat{g}_{\s{MMSE}} & = \underset{\hat{g} \in \s{SO}(3)} {\argmax} \int_{\s{SO(3)}} \langle \hat{g}, g \rangle \P_\Lambda(g|y, \barV)\der g \\ 
    & = \underset{\hat{g} \in \s{SO}(3)} {\argmax}  \langle \hat{g}, \hat{g}_{\s{relax}} \rangle. \label{eqn:g_MMSE_simlified}
\end{align}
As a consequence, \eqref{eqn:g_MMSE_simlified} is equal to the rotation $\hat{g} \in \s{SO}(3)$ that has the highest correlation with the relexed estimator $\hat{g}_{\s{relax}}$ as in \eqref{eqn:relaxEstimator}. It follows that
\begin{align}
    \hat{g}_{\s{MMSE}} = \underset{\hat{g} \in \s{SO}(3)} {\argmax}  \langle \hat{g}, \hat{g}_{\s{relax}} \rangle = \underset{\hat{g} \in \s{SO}(3)} {\argmin} \norm{\hat{g} - \hat{g}_{\s{relax}}}_F^2
\end{align}
Thus, $\hat{g}_{\s{MMSE}}$ is exactly the Procrustes procedure performed on the intermediate estimator $\hat{g}_{\s{relax}}$.

\section{Proof of Proposition \ref{thm:mapAndMmseEstimatorsConicideness}} \label{apx:mapAndMmseEstimatorsConicideness}

We split the proof into two parts. First, we show that 
\begin{align}
    \lim_{\sigma \to 0} \hat{g}_{\s{MAP}} = \hat{g}_{\s{MLE}}.\label{eq:MAP_MLE}
\end{align}
Second, we show that 
\begin{align}
    \lim_{\sigma \to 0} \hat{g}_{\s{MMSE}} = \hat{g}_{\s{MLE}}.\label{eq:MMSE_MLE}
\end{align}

For \eqref{eq:MAP_MLE}, the argument is straightforward. 
Since the prior density $\Lambda(g)$ is bounded below by some positive constant, 
the term $\log(\mathrm{d}\Lambda(g))$ becomes negligible as $\sigma \to 0$ compared to the other term. 
Consequently, the MAP estimator converges to the MLE in this limit.

For \eqref{eq:MMSE_MLE}, we apply the following theorem \cite[Corollary 5.11]{robert1999monte}:
\begin{thm} \label{thm:auxThm}
Consider $h$ a real-valued function defined on a closed and bounded set $\Theta$, of $\mathbb{R}^p$. Let $\pi$ be a positive density on $\Theta$. If there exists a unique solution $\theta^{*}$ satisfying 
\begin{align}
    \theta^{*} = \underset{\theta \in \Theta} {\argmax} \ h(\theta),
\end{align}
then
\begin{align}
     \underset{\lambda \to \infty} {\lim} \frac{\int_{\Theta} \theta e^{\lambda h(\theta)}\pi(\theta)\der \theta}{\int_{\Theta} e^{\lambda h(\theta)}\pi(\theta)\der \theta} = \theta^{*},
\end{align}
provided $h$ is continuous at $\theta^*$.
\end{thm} 
Let $h(g)=-\norm{y-\Pi( g \circ \barV)} ^2$ and $\pi(g)\der g=\der \Lambda(g)$. We immediately get 
\begin{align}
    \nonumber \lim_{\sigma \to 0} \hat{g}_{\mathrm{relax}} & = \lim_{\sigma \to 0} \frac{\int_{\s{SO}(3)} g \exp\Big(-\norm{y-\Pi (g \circ \barV)}^2 / 2\sigma^2\Big)\der \Lambda \p{g}}{\int_{\s{SO}(3)} \exp\Big(-\norm{y-\Pi (g \circ \barV)}^2 / 2\sigma^2\Big)\der \Lambda \p{g}} \\ \nonumber & =  \lim_{\lambda \to \infty}  \frac{\int_{\s{SO}(3)} g \exp\Big(-\lambda \norm{y-\Pi (g \circ \barV)}^2\Big)\der \Lambda(g)}{\int_{\s{SO}(3)} \exp\Big(-\lambda \norm{y-\Pi (g \circ \barV)}^2 \Big)\der \Lambda(g)} \\& = \underset{g \in \s{SO}(3)} {\argmax} \ h(g) = \hat{g}_{\s{MLE}}.
\end{align}
Finally, we note that $\hat{g}_{\s{MMSE}}$ is obtained by applying the Orthogonal Procrustes procedure to project $\hat{g}_{\s{relax}}$ back onto $\s{SO}(3)$. 
Since $\hat{g}_{\s{MLE}}$ is itself an element of $\s{SO}(3)$, it follows naturally that
\begin{align}
    \lim_{\sigma \to 0} \hat{g}_{\s{MMSE}} = \hat{g}_{\s{MLE}}.
\end{align}

\section{Non-uniform distribution of rotations}
\subsection{Isotropic Gaussian distribution of {SO}(3) rotations} \label{apx:isotropicGaussianDef}
The IG distribution $\mathcal{I}\mathcal{G}_{\s{SO}(3)}(\eta)$, parameterized by the scalar variance $\eta^2$, can be represented in an axis-angle form, with uniformly sampled axes and a rotation angle $\omega \in [0, \pi]$, having the following probability density function \cite{savjolova1985preface}:
\begin{align}
    f(\omega) = \frac{1-\cos \omega}{\pi} \sum_{\ell=0}^{\infty} (2\ell+1)e^{-\ell(\ell+1)\eta^2} \frac{\sin \left((\ell+ 1/2)\omega\right)}{\sin (\omega/2)}. \label{eqn:appB1}
\end{align}
Notably, the uniform distribution on $\s{SO}(3)$, denoted $\mathcal{U}_{\s{SO}(3)}$, corresponds to uniformly sampled axes and is characterized by the density function $f(\omega) = \frac{1 - \cos \omega}{\pi}$, which represents the limiting case of $\eta \to \infty$ in \eqref{eqn:appB1}.

To provide intuition for the spread induced by $\eta$, consider the concentrated regime $\eta\ll 1$. In this regime, one may use the standard small-angle approximation in which a rotation is represented by a rotation vector $r\in\mathbb{R}^3$ with
\begin{align}
    r \sim \mathcal{N}(0,\eta^2 I_3), \qquad \omega \approx \|r\| , \label{eqn:small_angle_approx}
\end{align}
so that $\omega$ is approximately Maxwell distributed with scale parameter $\eta$~\cite{johnson1995continuous}.
Consequently, its standard deviation is
\begin{align}
    \mathrm{Std}(\omega) \approx \eta\,\sqrt{3-\frac{8}{\pi}} = \eta\,\sqrt{\frac{3\pi-8}{\pi}} \approx 0.67\,\eta \quad \text{(radians)}.
    \label{eqn:std_omega_eta}
\end{align}
For example, $\eta=0.1$ corresponds to $\mathrm{Std}(\omega)\approx 0.067$ rad $\approx 3.8^\circ$.

For $\eta > 1$, the series converges quickly, and $\ell_{\text{max}} = 5$ is typically sufficient for sub-percent accuracy. However, as $\eta$ decreases, convergence becomes slower, making it inefficient for modeling concentrated distributions. Fortunately, this series has been well studied, and an excellent approximation for $\eta < 1$ has been proposed in the literature \cite{matthies1988normal}, given by the following closed-form approximation:
\begin{align}
    f(\omega) \simeq \frac{1-\cos \omega}{\sqrt{\pi}} \eta^{-\frac{3}{2}} e^{\left(\frac{\eta}{4} - \frac{\left(\omega/2\right)^2}{\eta}\right)} \left(\frac{\omega - e^{-\pi^2/\eta}\left(\left(\omega-2\pi\right)e^{\pi \omega / \eta} + \left(\omega+ 2\pi\right)e^{-\pi \omega / \eta}\right)}{2\sin\left(\frac{\omega}{2}\right)}\right).
\end{align}

The method for generating samples according to the isotropic Gaussian distribution on $\s{SO}(3)$ relies on the inverse sampling theorem. This technique enables sampling from a specified probability distribution by utilizing its cumulative distribution function (CDF). Specifically, it involves generating a uniform random number between 0 and 1 and then applying the inverse CDF to obtain a sample from the target distribution~\cite{savjolova1985preface}. 

\subsection{Estimating the rotational prior} \label{sec:prior-dist-est}
A key practical ingredient in Bayesian pose estimation is the prior over rotations. From an algorithmic standpoint, two complementary approaches are currently being developed to handle \emph{unknown} viewing-angle distributions. First, building on EM-type refinement, recent work proposes to introduce an explicit parameterization of the orientation distribution over $\mathsf{SO}(3)$ and to jointly optimize the structure and the viewing-angle distribution in the M-step, effectively recovering the rotational prior from the data \cite{xu2025misspecified}. This perspective generalizes Bayesian EM refinement pipelines such as \textsc{RELION} by treating the pose prior as an estimable object that can be learned jointly with the structure during refinement \cite{scheres2012relion}. Second, \emph{method-of-moments} approaches relax the uniformity assumption and aim to recover both the structure and the unknown viewing-angle distribution from low-order statistics, typically via alternating optimization between the two components \cite{sharon2020method,kileel2025two}.

In practice, we recommend starting with a uniform (or weak) prior when little is known about the rotational prior, and then optionally estimating a smoothed viewing-angle distribution from an initial refinement and plugging it into the MMSE estimator; simple sensitivity checks (uniform vs.\ learned priors, and varying regularization) can help diagnose the effect of preferred orientations. We emphasize, however, that principled estimation of the rotational prior, and a full theory of robustness to prior misspecification in cryo-EM/ET, remains an important open direction for future work.

\section{Proof of Proposition \ref{thm:relationBetweenMMSEandEM}} \label{apx:relationBetweenMMSEandEM}
Following \eqref{eqn:emSoftAssignmentUpadteRule}, the update law (M-step) of the $t+1$ iteration of the expectation-maximization algorithm is given by:
    \begin{align}
        \hat{V}^{(t+1)} =  \underset{V} {\argmin} \sum_{i=0}^{M-1} \int_{g\in\s{SO}(3)} p_{i}^{(t)}(g) \norm{y_i - \p{ g \circ V}}^2 \der g, \label{eqn:mStepApx}
    \end{align}
where 
    \begin{align}
        p_{i}^{(t)}(g) = \frac{\exp\Big(-\norm{y_i- \p{ g \circ \hat{V}^{(t)}}}^2/2\sigma^2\Big)}{ \int_{g\in\s{SO}(3)} \exp\Big(-\norm{y_i- \p{g \circ \hat{V}^{(t)}}}^2/2\sigma^2\Big)\der g}.
    \end{align}
As the action by the group element $g$ preserves the norm, we have,
    \begin{align}
        \norm{y_i - \p{ g \circ V}}^2 = \norm{g^{-1}\circ {y_i} - V}^2 = \norm{g^{-1} \circ y_i}^2 + \norm{V}^2 - 2 {\, \langle{g^{-1} \circ y_i}, V \rangle}. \label{eqn:normPreservation}
    \end{align}
Thus, substituting \eqref{eqn:normPreservation} into \eqref{eqn:mStepApx}, leads to:
    \begin{align}
        \hat{V}^{(t+1)} =  \underset{V} {\argmin} \sum_{i=0}^{M-1} \int_{g\in\s{SO}(3)} p_{i}^{(t)}(g) \ppp{\norm{g^{-1} \circ y_i}^2 + \norm{V}^2 - 2 {\, \langle{g^{-1} \circ y_i}, V \rangle}} \der g. \label{eqn:mStepNormPreseveApx}
    \end{align}
As the term $\norm{g \circ y_i}^2$ is independent of $V$, \eqref{eqn:mStepNormPreseveApx} can be simplified to:
    \begin{align}
        \hat{V}^{(t+1)} =  \underset{V} {\argmin} \sum_{i=0}^{M-1} \int_{g\in\s{SO}(3)} p_{i}^{(t)}(g) \ppp{\norm{V}^2 - 2 {\, \langle{g^{-1} \circ y_i}, V \rangle}} \der g. \label{eqn:mStepSimplifiedApx}
    \end{align}
Let us denote by $h(V)$ the right hand side of \eqref{eqn:mStepSimplifiedApx}:
    \begin{align}
        h(V) \triangleq \sum_{i=0}^{M-1} \int_{g\in\s{SO}(3)} p_{i}^{(t)}(g) \ppp{\norm{V}^2 - 2 {\, \langle{g^{-1} \circ y_i}, V \rangle}} \der g.
    \end{align}
Taking the first-order condition with respect to $V$ gives 
    \begin{align}
        \frac{\partial}{\partial V} h(V) = 2 \sum_{i=0}^{M-1} \int_{g\in\s{SO}(3)} p_{i}^{(t)}(g) \p{V - g^{-1} \circ y_i} \der g = 0. \label{eqn:mStepDerivative}
    \end{align}
As $\int_{g\in\s{SO}(3)}p_{i}^{(t)}(g) = 1$ for every $i$, we can simplify \eqref{eqn:mStepDerivative} to obtain:
    \begin{align}
        V =& \frac{1}{M} \sum_{i=0}^{M-1} \int_{g\in\s{SO}(3)} p_{i}^{(t)}(g)\cdot (g^{-1}\circ y_i)  \notag \\
        =&  \frac{1}{M} \sum_{i=0}^{M-1} \big(\int_{g\in\s{SO}(3)} p_{i}^{(t)}(g)\cdot g^{-1} \der g \big)\circ y_i. \label{eqn:c7}
    \end{align}
Thus, using the definition \eqref{eqn:g_MMSE_operator_action} we obtain 
\begin{align}
    \hat{V}^{(t+1)} = \frac{1}{M} \sum_{i=0}^{M-1} \hat{\mathfrak{g}}_{\s{MMSE}, i,t} \circ y_i, 
\end{align}
which proves the proposition.

\end{appendices}

\end{document}